\documentclass[numberedappendix]{emulateapj}
\usepackage{longtable}

\shorttitle{The Physical Properties and Kinematics of Molecular Gas in the Antennae Galaxies}
\shortauthors{Ueda et al.}

\begin{document}

\title{Unveiling the physical properties and kinematics 
of molecular gas in the Antennae Galaxies (NGC~4038/9) 
through high resolution CO~($J$ = 3--2) observations}
\author{
Junko Ueda\altaffilmark{1,2}, 
Daisuke Iono\altaffilmark{1,3}, 
Glen Petitpas\altaffilmark{3}, 
Min S. Yun\altaffilmark{4}, 
Paul T. P. Ho\altaffilmark{3,5}, 
Ryohei Kawabe\altaffilmark{1}, 
Rui-Qing Mao\altaffilmark{6}, 
Sergio Mart\'in\altaffilmark{7}, 
Satoki Matsushita\altaffilmark{5,8}, 
Alison B. Peck\altaffilmark{3,8}, 
Yoichi Tamura\altaffilmark{1,9}, 
Junzhi Wang\altaffilmark{10}, 
Zhong Wang\altaffilmark{3}, 
Christine D. Wilson\altaffilmark{11}, 
and Qizhou Zhang\altaffilmark{3}
}
\altaffiltext{1}{Nobeyama Radio Observatory, NAOJ, 462-2 Minamimaki, Minamisaku, Nagano, 384-1305, Japan}
\altaffiltext{2}{The University of Tokyo, 7-3-1 Hongo, Bunkyo-ku, Tokyo 133-0033, Japan}
\altaffiltext{3}{Harvard-Smithsonian Center for Astrophysics, 60 Garden Street, Cambridge, MA 02138, USA}
\altaffiltext{4}{Department of Astronomy, University of Massachusetts, Amherst, MA 01003, USA}
\altaffiltext{5}{Institute of Astronomy and Astrophysics, Academia Sinica, P.O. Box 23-141, Taipei 10617, Taiwan, R.O.C.}
\altaffiltext{6}{Purple Mountain Observatory, Chinese Academy of Sciences, 210008 Nanjing. China}
\altaffiltext{7}{European Southern Observatory, Alonso de C\'ordova 3107, Vitacura, Casilla 19001, Santiago 19, Chile}
\altaffiltext{8}{Joint ALMA Observatory, Alonso de C\'ordova 3107, Vitacura, Casilla 19001, Santiago 19, Chile}
\altaffiltext{9}{Institute of Astronomy, The University of Tokyo, 2-21-1 Osawa, Mitaka,Tokyo, 181-0015, Japan}
\altaffiltext{10}{Department of Astronomy, Nanjing University, 22  Road, Nanjing 210093, China}
\altaffiltext{11}{Department of Physics and Astronomy, McMaster University, Hamilton, ON L8S 4M1, Canada}

\begin{abstract}
We present a $\sim 1\arcsec$~(100~pc) resolution $^{12}$CO~(3--2) map 
of  the nearby intermediate stage interacting galaxy pair NGC~4038/9 
(the Antennae galaxies) obtained with the Submillimeter Array.  
We find that half the CO~(3--2) emission originates in the overlap region 
where most of the tidally induced star formation had been previously found 
in shorter wavelength images, with the rest being centered on each of the nuclei.  
The gross distribution is consistent with lower resolution single dish images, 
but we show for the first time the detailed distribution 
of the warm and dense molecular gas across this galaxy pair at resolutions 
comparable to the size of a typical giant molecular complex.  
While we find that 58\% (33/57) of the spatially resolved Giant Molecular Associations 
(GMAs;  a few $\times$ 100 pc) are located in the overlap region, 
only $\leq 30\%$ spatially coincides with the optically detected star clusters, 
suggesting that the bulk of the CO~(3--2) emission traces the regions 
with very recent or near future star formation activity.  
The spatial distribution of the CO~(3--2)/CO~(1--0) integrated brightness temperature ratios 
mainly range between 0.3 and 0.8, which suggests that 
on average the CO~(3--2) line in the Antennae is not completely thermalized 
and similar to the average values of nearby spirals.  
A higher ratio is seen in both nuclei and the southern complexes in the overlap region.  
Higher radiation field associated with intense star formation can account for 
the nucleus of NGC 4038 and the overlap region, but the nuclear region of NGC 4039 
show relatively little star formation or AGN activities and cannot be easily explained.  
We show kinematical evidence that the high line ratio in NGC~4039 
is possibly caused by gas inflow into the counter-rotating central disk.  
\end{abstract}

\keywords{galaxies: individual (NGC~4038, NGC~4039) --- galaxies: interactions --- galaxies: ISM --- radio lines: galaxies}

\section{Introduction} 
Gravitational interaction between galaxies is now considered 
to be one of the primary mechanisms of galaxy evolution.  
Major mergers, composed of two or more galaxies with comparable masses, 
are widely believed to provide a way to form elliptical and spheroidal galaxies 
\citep{Toomre77}.  
Furthermore, galaxy interaction is known to increase star formation activities 
as seen in the increasing fraction of tidally distorted morphologies 
in ultraluminous infrared galaxies \citep[ULIRGs: L$_{\rm FIR} \geq 10^{12}$ L$_{\sun}$, 
SFR $\sim 10^{2}$ M$_{\sun}$ yr$^{-1}$;][]{Sanders96}.  
In addition, the number of merging galaxies appears to increase at high redshifts 
\citep[e.g.][]{Bundy09}.
The obvious first step in characterizing this process and the response of the cold gas 
and its relation to merger induced starburst activity is to observe nearby merging galaxies.  

The Antennae galaxies (NGC~4038/9) is the nearest colliding galaxy pair observable 
from the northern hemisphere located at a distance of 22 Mpc \citep{Schweizer08}.  
Numerical simulations of the Antennae \citep[e.g.][]{Toomre72} 
suggested that the Antennae are at the intermediate stage of the collision.  
\citet{Mihos93} suggest that the two spiral galaxies passed the initial pericenter 
210~Myr ago, and the final coalescence will take place in about 100~Myr.  

The Antennae have been observed at all available wavelengths from radio to X-ray.  
\citet{Whitmore99} observed the Antennae with the Wide Field Planetary Camera 2 
on the \textit{Hubble Space Telescope}~(HST) and identified thousands of stellar clusters 
which have formed during the merging process.  
They found that most of the star formation occurs in the form 
of massive compact star clusters, which are formed from supergiant molecular clouds.  
Previous observations have found that most of the active star forming regions are 
concentrated in the region where the two galaxies overlap.  
The \textit{Herschel}-PACS maps at 70, 100 and 160 $\micron$ revealed 
that both nuclei are brighter than the HII regions in the arms, 
but the brightest emission in three bands comes from the overlap region \citep{Klaas10}.  
The mid-infrared (MIR) and far-infrared (FIR) emission traces buried star formation, 
which is obscured by dust and therefore is invisible at optical wavelengths.  
They estimated that the integral SFR is $\sim$ 22 M$_{\sun}$ yr$^{-1}$ 
and the SFR in the southernmost knot of the overlap region 
with an area of 4.68 kpc$^{2}$ is $\sim$ 3 M$_{\sun}$ yr$^{-1}$.  
They also found that several emission knots in the overlap region 
are in different evolutionary stages.  
Low angular resolution observations have revealed strong [CII] emission,
which arises mainly from photo dissociation regions created by far-ultraviolet photons 
from hot young stars, near the overlap region \citep{Nikola98}.  
In addition, the southernmost knot of the overlap region is 
currently the most active area with a very young stellar complex.  
\citet{Neff00} observed 4 and 6 cm radio continuum emission 
using the VLA and estimated that the overall star formation rate 
seems to be a factor of $\sim$ 20 higher than normal galaxies.  
They found that the compact radio source with the strongest thermal radio emission 
requires $\sim$ 5000 O5 stars to account for the free-free emission.  
They estimate a supernova rate of $\sim$ 0.2 yr$^{-1}$ 
if the sources with steep radio spectra are dominated 
by non-thermal emission produced in supernova remnants.  

The distribution and the characteristics of molecular gas 
have been investigated in detail using mm/submm telescopes around the world.  
\citet{Gao01} obtained the CO~(1--0) map using the NRAO 12~m single-dish telescope 
and found that the distribution of the CO~(1--0) emission is similar to 
those of the MIR and FIR emission at a kiloparsec scale.  
\citet{Wilson00} identified five supergiant molecular complexes (SGMCs) 
with masses of (3--6) $\times$ 10$^{8}$ M$_{\sun}$ in the overlap region 
using the $^{12}$CO~(1--0) map obtained at the Owens Valley Radio Observatory (OVRO).    
\citet{Zhu03} have obtained single-dish maps at the Nobeyama 45m telescope 
in the CO~(1--0) line and using the James Clark Maxwel Telescope (JCMT) 
in the CO~(2--1) and CO~(3--2) lines.  
Each nucleus contains a molecular mass of $\sim 10^{9}$ M$_{\sun}$ 
and the overlap region accounts for approximately 42$\%$ of total $^{12}$CO~(1--0) 
flux of the inner disk of the Antennae.  
They found that the CO integrated intensity ratios are high in the overlap region.  
This is the first case where high CO~(3--2)/CO~(1--0) ratios 
are found outside a galactic nucleus.  
Furthermore, \citet{Schulz07} obtained CO~(1--0) and CO~(2--1) maps 
using the IRAM 30~m Millimeter Radio Telescope 
and CO~(3--2) map using the Heinrich Hertz Telescope.  
The total molecular gas mass of the system ($\sim$ 10$^{10}$ M$_{\sun}$) is 
about twice the total gas mass of the Milky Way.  
However the molecular gas mass within the 1 kpc nucleus of NGC~4038 and NGC~4039 
exceeds that of the central 1 kpc of the Milky Way by a factor of almost 100,   
revealing high gas concentration into the two nucleus.  
\citet{Schulz07} also derived the line ratios ranging between 0.50 and 0.66, 
which are lower than the ratios estimated by \citet{Zhu03}.  

The purpose of our work is to investigate the physical properties and the kinematics 
of the CO~(3--2) emitting molecular complexes in the Antennae 
using $\sim 1^{\prime\prime}$ (1$^{\prime\prime}$ corresponds to about 107 pc) 
resolution CO~(3--2) observations.  
The CO~(1--0) rotational transition has a critical density 
of $n_{\rm crit} \sim 10^{2.5}$ cm$^{_3}$ 
and an upper energy level $E_{\rm u} \sim$ 5~K, 
whereas the CO~(3--2) rotational transition has 
$n_{\rm crit} \sim 10^{4}$ cm$^{_3}$ and $E_{\rm u} \sim$ 33 K.  
Thus the CO~(3--2) emission is a better tracer 
of denser and warmer molecular gas than CO~(1--0) emission.  

This paper is organized as follows.  
We describe our observations in \S 2 and our results in \S 3.  
In \S 4, we provide a discussion of our results.  
We present the properties and the distribution 
of identified molecular complexes (\S 4.1), 
the CO~(3--2)/(1--0) brightness temperature ratio (\S 4.2),
the kinematics in NGC~4039 (\S 4.3), 
and a possible molecular gas bubble in NGC~4038 (\S 4.4).  
We summarize and conclude this paper in \S 5.

\section{Observations} 
We conducted CO~(3--2) observations of main bodies of the NGC~4038/9 
with the Submillimeter Array\footnote{
The Submillimeter Array is a joint project between the Smithsonian Astrophysical 
Observatory and the Academia Sinica Institute of Astronomy and Astrophysics 
and is funded by the Smithsonian Institution and the Academia Sinica.  
} \citep[SMA;][]{Ho04} 
on March 2005 and March 2008 in the compact configuration 
and December 2008 in the extended configuration.  
The data were obtained using a five-pointing mosaic 
where the phase centers were chosen to cover the main CO~(3--2) emission peaks 
seen in the JCMT map of \citet{Zhu03}.  
The coordinates of the phase centers are shown in Table \ref{tb:fov} 
and the fields of view are shown in Figure \ref{fig:IM} (right).  
The total on source observing time was approximately 10 hours.  
The SIS receivers were tuned to the frequency 
of the CO~(3--2) line ($\nu_{\rm rest}$=345.796 GHz) 
redshifted to the systemic velocity of 1634 km/s.  
The primary beam of the array at this frequency is $32{\arcsec}$.  
The correlator had a 2 GHz total bandwidth 
with a 0.8 MHz (0.7 km s$^{-1}$) frequency resolution.  
The quasar 3C 279 was observed for phase and amplitude calibration.  
Absolute flux calibration was performed using Uranus.  
The uncertainty of flux calibration is 20 $\%$.  

Data inspection and calibration was carried out using the MIR package 
written in IDL and imaging was done using the MIRIAD package.  
The synthesized beam size is $1\farcs 42 \times 1\farcs 12$, 
which corresponds to 150 pc $\times$ 120 pc  
with a position angle of 29.4$\arcdeg$.  
Natural weighting of the visibilities is used to maximize the sensitivity, 
and the achieved rms noise level in 10 km s$^{-1}$ channel maps 
is 38 mJy beam$^{-1}$.  
The continuum emission is not detected 
with an upper limit of 17.7 mJy (3 $\sigma$).

\section{Results} 
\subsection{CO~(3--2) Distribution and Total Flux} 
\begin{figure}[htbp]
 \begin{minipage}{0.5\hsize}
  \begin{center}
   \includegraphics[width=105mm,clip,trim=32 0 0 35]{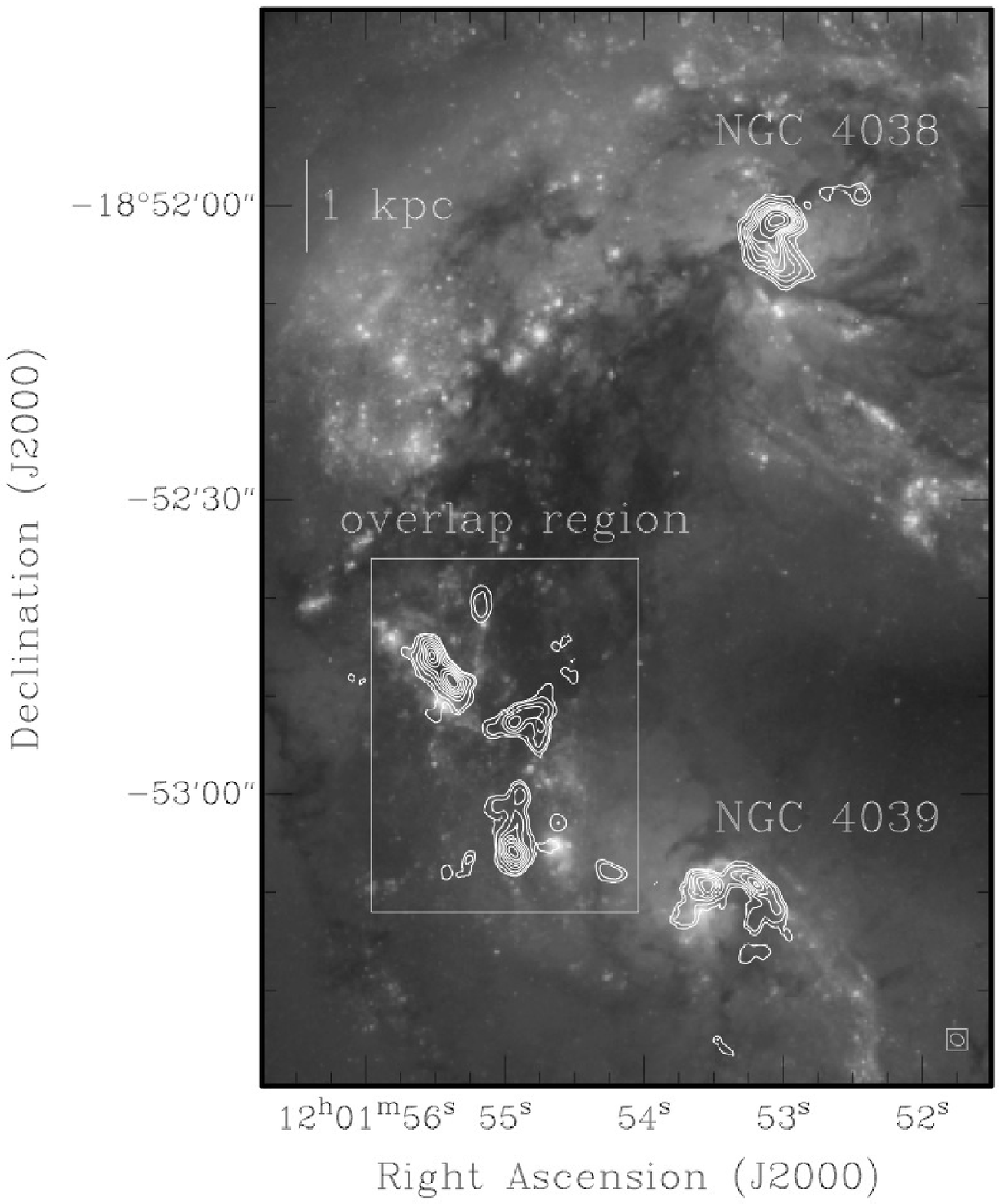}
  \end{center}
 \end{minipage}
 \begin{minipage}{0.5\hsize}
  \begin{center}
   \includegraphics[width=97mm]{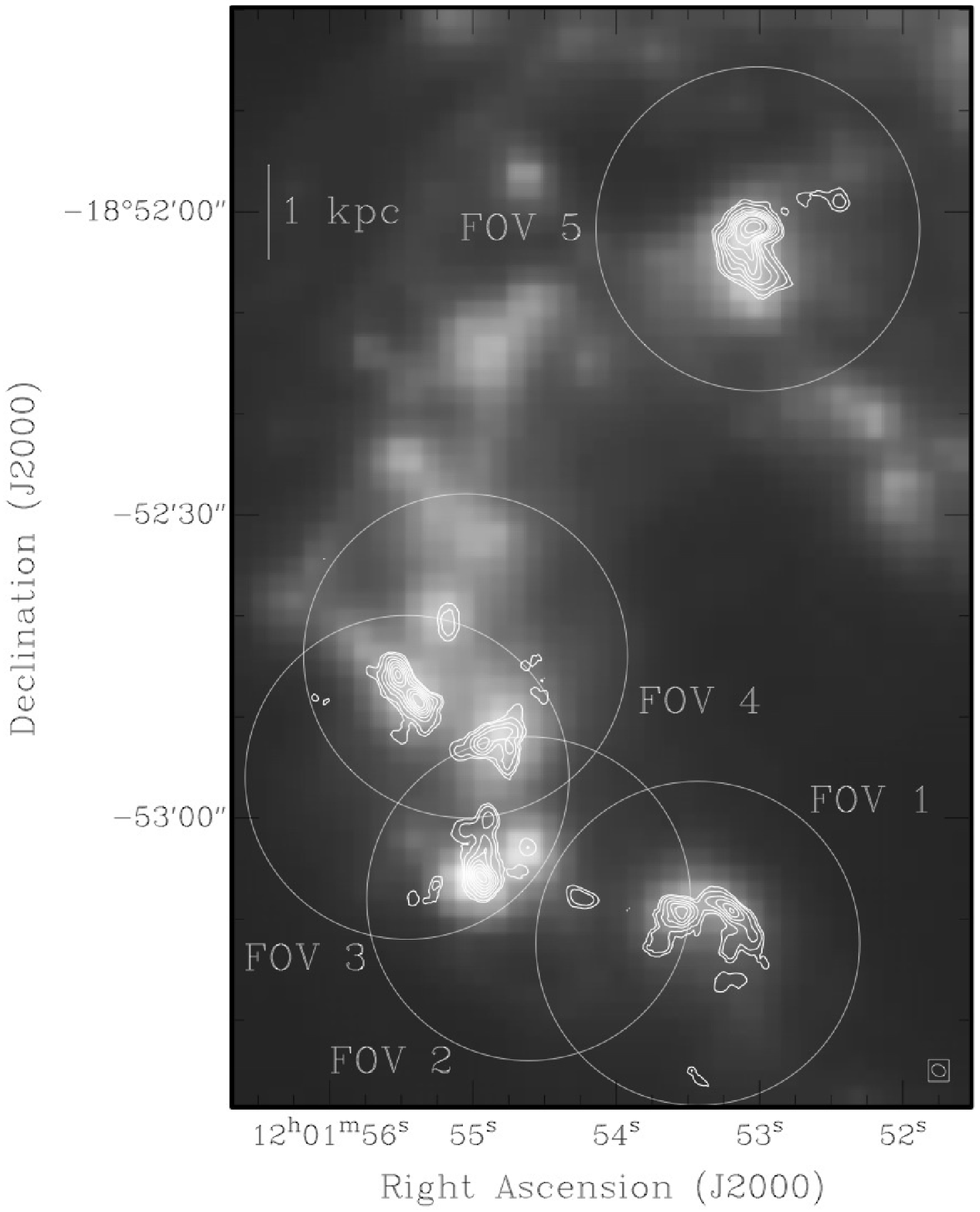}
  \end{center}
 \end{minipage}
 \caption{
	The CO~(3--2) integrated intensity (between 1304 and 1804~km~s$^{-1}$) contour map 
	overlaid on the HST 435 nm image (left) and the Spitzer 8 $\mu$m image (right).  
	The contour levels are 6~Jy km s$^{-1}$$\times$ 2, 3, 5, 7, 9, 11, 13, 15, 20, 25.  
	The galaxy and region names are labeled in the left figure.  
	The circles in the right figure show the primary beam centered on each pointing.  
	The ellipse in the lower right corner of both figures shows the synthesized beam size  
	($1\farcs42 \times 1\farcs12$).  
	These integrated intensity maps are clipped at 2$\sigma$ level using the AIPS task, MOMNT.  	 
 } 
 \label{fig:IM}
\end{figure}

The CO~(3--2) integrated intensity map overlaid on optical/MIR images 
is shown in Figure \ref{fig:IM}.  
In addition, we compare the CO~(3--2) map with multiwavelength images 
from radio and to X-ray (see Appendix A).  
In general, the locations of the CO~(3--2) emitting molecular complexes 
are consistent with the positions of the SGMCs detected by \citet{Wilson00} 
in CO~(1--0), but the higher angular resolution achieved 
in the CO~(3--2) map shows details that were unseen in the CO~(1--0) image.  
Both galaxies show central CO~(3--2) concentrations near the nuclei 
and have one gas arm roughly distributed along the dust lanes toward southwest 
seen in the HST 435 nm image (Figure \ref{fig:IM} (left)).  
The strongest CO~(3--2) emission is associated with 
the nucleus of NGC~4038 with a peak flux density 
(10 km s$^{-1}$ velocity resolution) of 2.0 $\pm$ 0.4 Jy beam$^{-1}$.  
The CO~(3--2) emission near the nucleus of NGC~4039 is composed 
of at least two distinct large scale components separated by 560~pc.  
The overlap region consists of numerous emission peaks, 
with complex structure seen in larger ($\sim$ 1 kpc) molecular complexes. 
Spatial correlation with optical emission is poor in most of 
the molecular complexes in the overlap region, 
mostly because these regions are strongly affected by dust extinction.  
However, the general association with the \textit{Spitzer} 8 $\mu$m emission 
(Figure \ref{fig:IM} (right)), which traces the dusty regions associated with 
CO~(3--2) emitting molecular clouds, 
is much better than the optical association.  

The total integrated CO~(3--2) flux of main body of the Antennae is 
4.4 $\times$ 10$^{3}$ Jy km s$^{-1}$.  
The two nuclei of NGC~4038 and NGC~4039 contain 1.4 $\times$ 10$^{3}$ 
and 9.4 $\times$ 10$^{2}$ Jy km s$^{-1}$, respectively.  
The total flux in the overlap region is 2.1 $\times$ 10$^{3}$ Jy km s$^{-1}$, 
which is approximately half of the total integrated flux over the entire galaxy.  
The percentage of missing flux derived by comparing the SMA 
and JCMT flux \citep{Zhu03} are 55$\%$ in NGC~4038, 
35$\%$ in NGC~4039 and 67$\%$ in the overlap region.  
The low missing flux in NGC~4039 suggests the compact nature 
of the CO~(3--2) emitting clouds, while the relatively large missing flux 
suggests the presence of a significant extended ($\gtrsim$ 3 kpc) 
component in the overlap region.  

The central $\sim 100$~pc in NGC~4038 is unresolved 
by the $\sim 1 \arcsec$ beam, 
but the strong signal allows us to use the extended configuration data alone 
for a higher angular resolution image (Figure \ref{fig:extended}).  
The synthesized beam size is $0\farcs 82 \times 0\farcs 56$ 
by adopting natural weighting of the visibilities.  
Most of the diffuse and extended emission below 15$\sigma$ in the $\sim 1 \arcsec$ 
resolution map made using all data (Figure \ref{fig:IM}) is resolved out 
in the high resolution map.  
The deconvolved size of the main component is 
230~pc ($2.1 \arcsec$) $\times$ 150~pc ($1.4 \arcsec$), 
and the FWHM velocity width is 110~km~s$^{-1}$.  
The brightness temperature $T_{B}$ (the peak intensity) scales 
with the beam filling factor $f$ and the molecular gas temperature $T$.  
From the peak intensity of 13~K and 
assuming that the typical molecular gas temperature $T$ of the CO~(3--2) 
emission is $\lesssim$ 30~K \citep{Schulz07}, the beam filling factor is $\lesssim$ 0.4.  
Thus the distribution of the CO~(3--2) emission is likely 
more compact than the region observed with the $\sim75$~pc beam.

\subsection{CO~(3--2) Kinematics} 
The velocity fields of NGC~4038, NGC~4039, and the overlap region 
are shown in Figure~\ref{fig:mom1}~(a), (b) and (c), respectively.  
At this spatial resolution, the velocity at the K$_{s}$-band nucleus of NGC~4038 
\citep{Mengel02} is $\sim$ 80 km s$^{-1}$ lower than the surrounding medium, 
which may suggest a rotation or gas streaming into the nuclear region along the arm.  
In contrast, the CO~(3--2) emitting gas near the nucleus of NGC~4039 
shows a large velocity gradient (310~km~s$^{-1}$~kpc$^{-1}$) 
from the north-west to south-east, with the velocity ranging 
from 1560~km~s$^{-1}$ to 1640~km~s$^{-1}$.  
Furthermore, the gas component distributed along the optical dust lane 
of NGC~4039 also shows a steep velocity gradient of 170~km~s$^{-1}$~kpc$^{-1}$.  
The implications of these kinematic features will be discussed in \S 4.3.  
In the overlap region, the velocity of the gas ranges 
from V$_{\rm LSR}$ = 1300 to 1650 km s$^{-1}$.  
The large scale kinematics in the overlap region show 
a north-south velocity gradient, which is  consistent with the velocity field 
seen in the lower resolution JCMT map \citep{Zhu03}.  
The new SMA map reveals that each molecular component shows 
complex kinematics, with velocity gradients in excess of 100~km~s$^{-1}$~kpc$^{-1}$ 
in each molecular complex, some as large as 300~km~s$^{-1}$~kpc$^{-1}$.  

\begin{figure}[htbp]
 \begin{center}
  \includegraphics[width=120mm, angle=270]{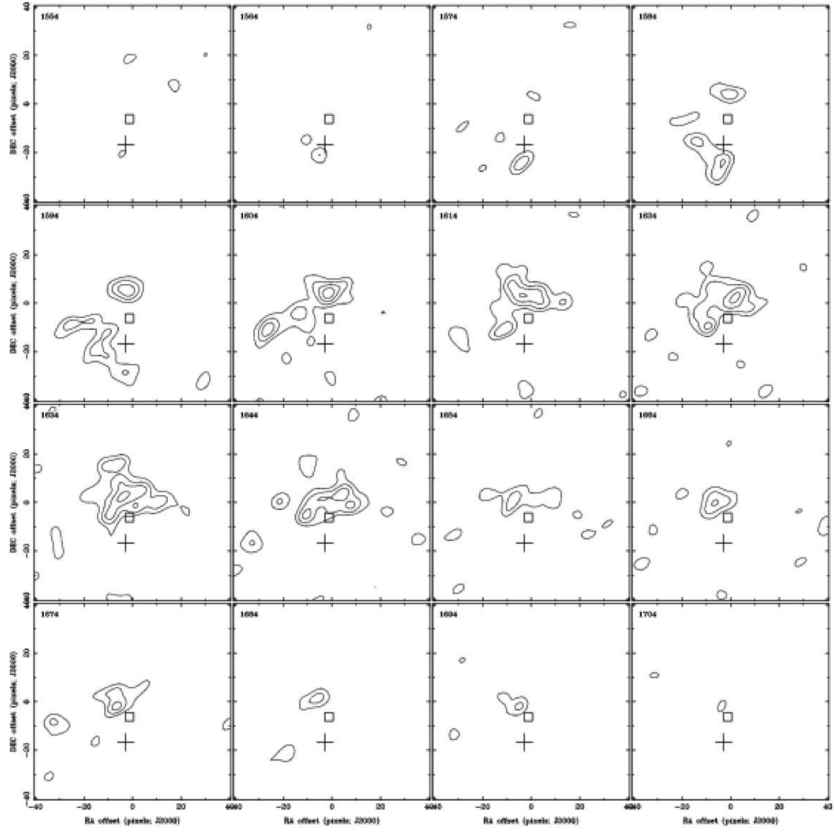}
 \end{center}
 \caption{
	The channel maps in NGC~4038 using only the data observed 
	in the extended configuration.  The contour levels are noise level 
	(1$\sigma$ = 70~mJy beam$^{-1}$) $\times$ 3, 4, 5, 6, 7, 8, 9, 10.  
	The cross sign shows the galactic center and the box sign shows 
	the radio source.  
 } 
 \label{fig:cmap_ex}
\end{figure}

The channel maps of the high resolution cube of NGC~4038 
with 10 km/s velocity resolution are shown in Figure \ref{fig:cmap_ex}.  
The rms noise level (1$\sigma$) is 70 mJy beam$^{-1}$.  
At this angular and velocity resolution, the morphology of the emission feature 
is highly complex around the nucleus, with the emission distributed 
in a semicircle around the known nonthermal radio source
dominated by synchrotron radiation from supernova remnants \citep{Neff00}.  
The radio source does not have detectable CO~(3--2) 
associated at an rms of 70 mJy beam$^{-1}$ 
which corresponds to a molecular gas mass limit of 9 $\times$ 10$^{5}$ M$_{\sun}$.  
Further discussion on the possible origin of this CO~(3--2) arc is provided in \S4.4.

\section{Discussion} 
\subsection{Properties of the CO~(3--2) Molecular Complexes} 
We identified molecular complexes by applying the automatic clump 
identification algorithm Clumpfind \citep{Williams94} to the CO~(3--2) data.  
One of programs in Clumpfind, CLFIND, works on 3D (RA, Dec, velocity) data cubes, 
searches for local peaks of emission, and follows them down to lower intensity levels.  
Then it decomposes the data cube into clumps in which the emission is concentrated.  
While the 2$\sigma$ threshold is the recommended value for identifying 
robust molecular clumps \citep{Williams94}, our experiments find that 
the total number of identified molecular complexes varies 
by 15$\%$ for a 10$\%$ deviation from  the 2$\sigma$ threshold.  
This is likely because the program falsely detects ($\sim 20\%$) 
sidelobes of the real emission.  
We find that the number of identified clumps only changes by 
a few percent for a threshold of $2.2 \leq \sigma \leq 2.6$.  
Thus we chose 2.6$\sigma$ for robustly detecting molecular clumps 
with high significance in the following analysis.  

A total of 57 molecular complexes 
are identified, 33 ($58\%$) of which are located in the overlap region.  
The properties of the molecular complexes such as position, systemic velocity ($V$), 
radiusv($r$), and velocity dispersion ($\sigma_{v}$) were measured 
using the program CLSTATS \citep{Williams94} and are summarized 
in Table \ref{tb:clump} (Appendix B).  
We estimate the error in the position, radius, and velocity to be 
$< 0\farcs1$, 44 pc, and 5~km~s$^{-1}$, respectively.  
The radii of the molecular complexes ranges between 85 and 348 pc, 
whereas the velocity dispersion ranges between 6 and 36 km s$^{-1}$.

\subsubsection{Molecular Mass} 
The molecular gas mass is derived by; 
\begin{eqnarray}
\frac{M_{CO}}{ \rm M_{\sun}} &=& 3.25 \times 10^{7}~X_{CO} \nonumber \\
&\times&\Bigl(\frac{S_{CO}~\Delta v}{\rm Jy~km~s^{-1}}\Bigr)\Bigl(\frac{\nu_{obs}}{\rm GHz}\Bigr)^{-2}\Bigl(\frac{D_{L}}{\rm Mpc}\Bigr)^{2}~(1+z)^{-3},
\end{eqnarray}
where $X_{CO}$ is the CO luminosity-to-H$_{2}$ mass conversion factor, 
$S_{CO}\Delta v$ is the velocity integrated flux, 
$\nu_{obs}$ is the observing frequency, 
and $D_{L}$ is the luminosity distance \citep{Solomon05}.  
We used $X_{CO}$ = 4.8 M$_{\sun}$/K~km~s$^{-1}$~pc$^{2}$ \citep{Solomon91} 
scaled by the average CO~(3--2)/(1--0) line intensity ratio for each region (see \S 4.2).  
The molecular masses are estimated to be 10$^{7}$ -- 10$^{8}$ M$_{\sun}$.  
In addition, the virial mass of the molecular complexes are estimated from 
$M_{vir} = 2rv^{2}/G$, where $r$ and $v$ is the radius and the velocity dispersion 
of the molecular complex, and $G$ is the gravitational constant.  
The molecular complex is gravitationally bound 
when their $M_{vir}/M_{CO}$ is smaller than unity.  
We find that 84~$\%$ of the molecular complexes are gravitationally bound, 
including the most massive molecular complex 
located north of the NGC~4038 nucleus.  
We note that the velocity dispersions used here are likely overestimated by a factor of 
2 -- 3 (\S 4.1.3), and hence the derived virial masses are also significantly overestimated.
This suggests that almost all of the molecular complexes observed at this spatial resolution 
have $M_{vir}/M_{CO} <$ 1 and that they are gravitationally bound.

\subsubsection{Radius and Velocity Dispersion} 
\begin{figure}[htbp]
 \begin{minipage}{0.3\hsize}
  \begin{center}
   \includegraphics[width=60mm]{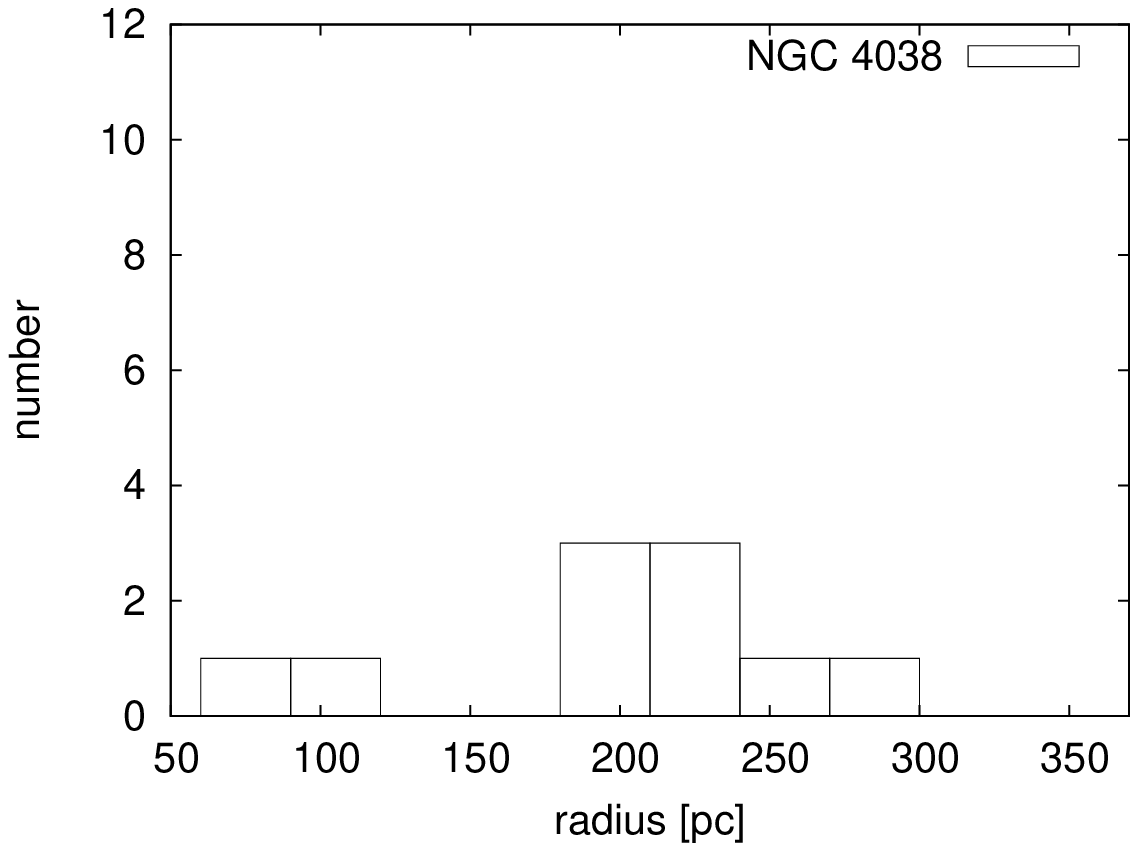}
  \end{center}
 \end{minipage}
 \begin{minipage}{0.3\hsize}
  \begin{center}
   \includegraphics[width=60mm]{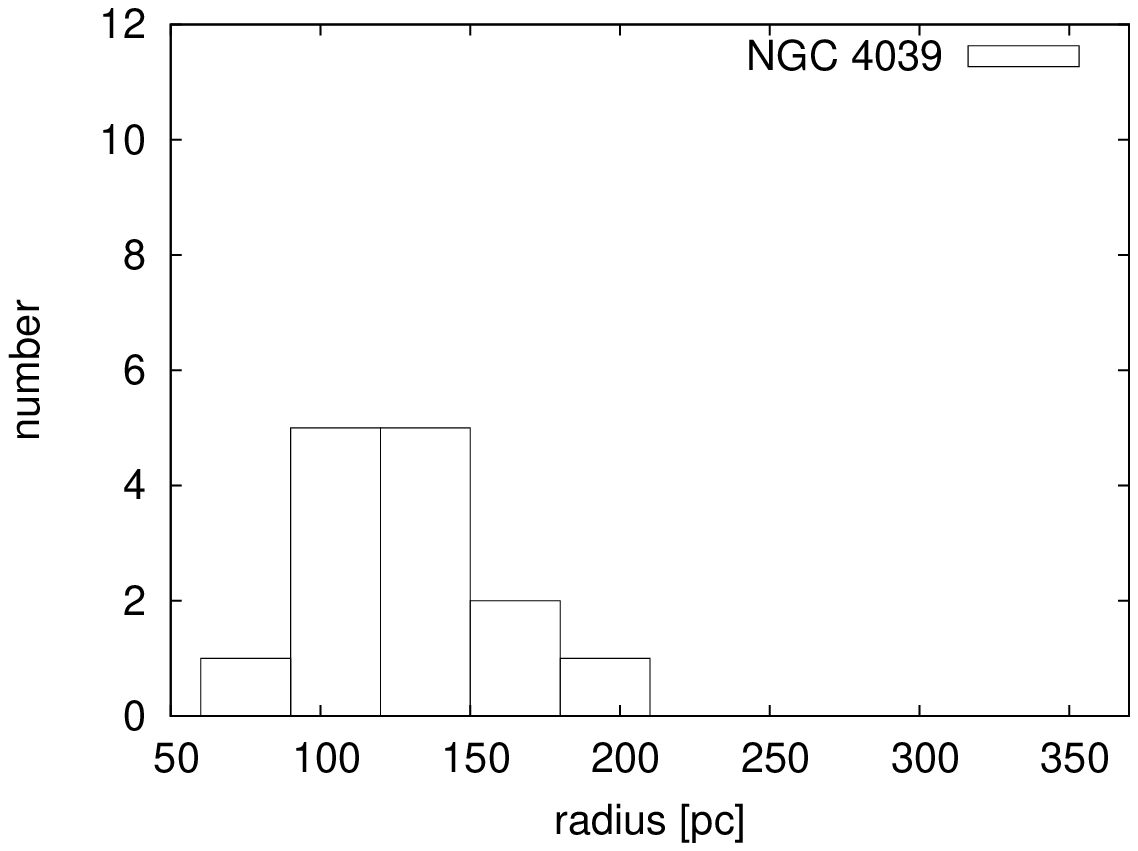}
  \end{center}
 \end{minipage}
 \begin{minipage}{0.3\hsize}
  \begin{center}
   \includegraphics[width=60mm]{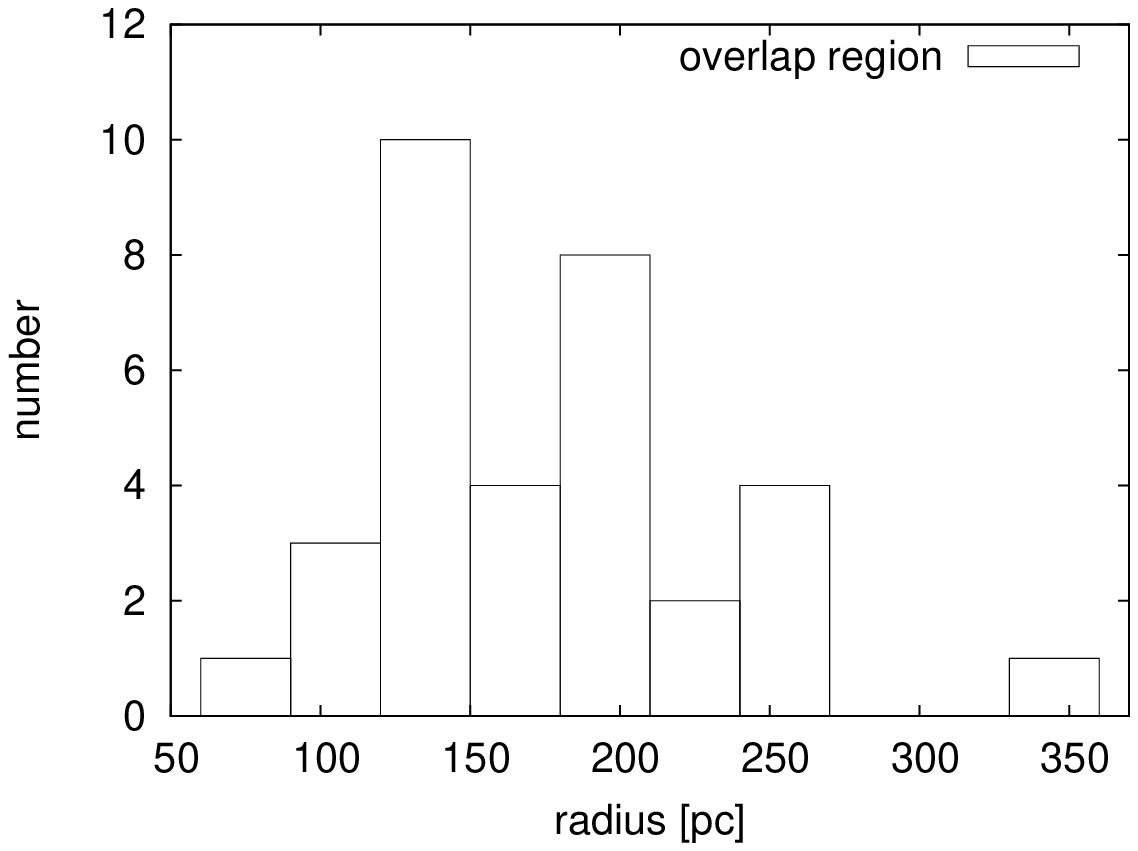}
  \end{center}
 \end{minipage}
 \caption{
	Distribution of the radii of the detected molecular complexes 
	in NGC~4038 (left), NGC~4039 (middle) and overlap region (right).  
 }
 \label{fig:r_hist}
\end{figure}

The distribution of the radii of the molecular complexes in each region 
are shown in Figure \ref{fig:r_hist}.  
The Kolmogorov-Smirnov test gave P-value = 0.002 (Table~\ref{tb:P-value}) 
for the comparison of the distribution between NGC~4038 and NGC~4039, 
suggesting that the two populations are significantly different.  
There are molecular complexes with large radii ($>$ 180 pc ) 
in NGC 4038, but not in NGC 4039.  
The median radii in NGC~4038 and NGC~4039 are 210 and 120 pc, respectively, 
and thus the molecular complexes near the nucleus of NGC~4038 are on average 
1.8 $\pm$ 0.7 times larger than those in NGC~4039.  
The size distribution in the overlap region range between 90 and 350 pc.  
The average size of the molecular complexes in the southern part of the overlap region 
(declination $<$ -18$^{\circ}$ 53${\arcmin}$ 00${\arcsec}$) 
is 1.3 times larger than the average size of those in the overlap region.  
In addition, the P-value = 0.975 and 0.017 (Table~\ref{tb:P-value}) 
for the comparison of the distribution between the southern part of the overlap region 
and NGC~4038, and NGC~4039, respectively.  
This suggests that the radii between NGC~4038 and 
the southern part of the overlap region are similar.  
\citet{Klaas10} estimated the SFRs of seven individual emission knots 
from their \textit{Herschel} maps.  
Knot K1, which corresponds to the southern part of the overlap region, 
has the highest SFR (2.78 M$_{\sun}$ yr$^{-1}$) and knot NN, 
which corresponds to the central region of NGC~4038, 
has the second highest SFR (1.17 M$_{\sun} $yr$^{-1}$).  
Therefore the larger molecular complexes, which are possibly formed 
by clustering around local peaks of the gravitational potential wells, 
may be used as the signposts of current dusty star forming regions.  

\begin{figure}[htbp]
 \begin{minipage}{0.3\hsize}
  \begin{center}
   \includegraphics[width=60mm]{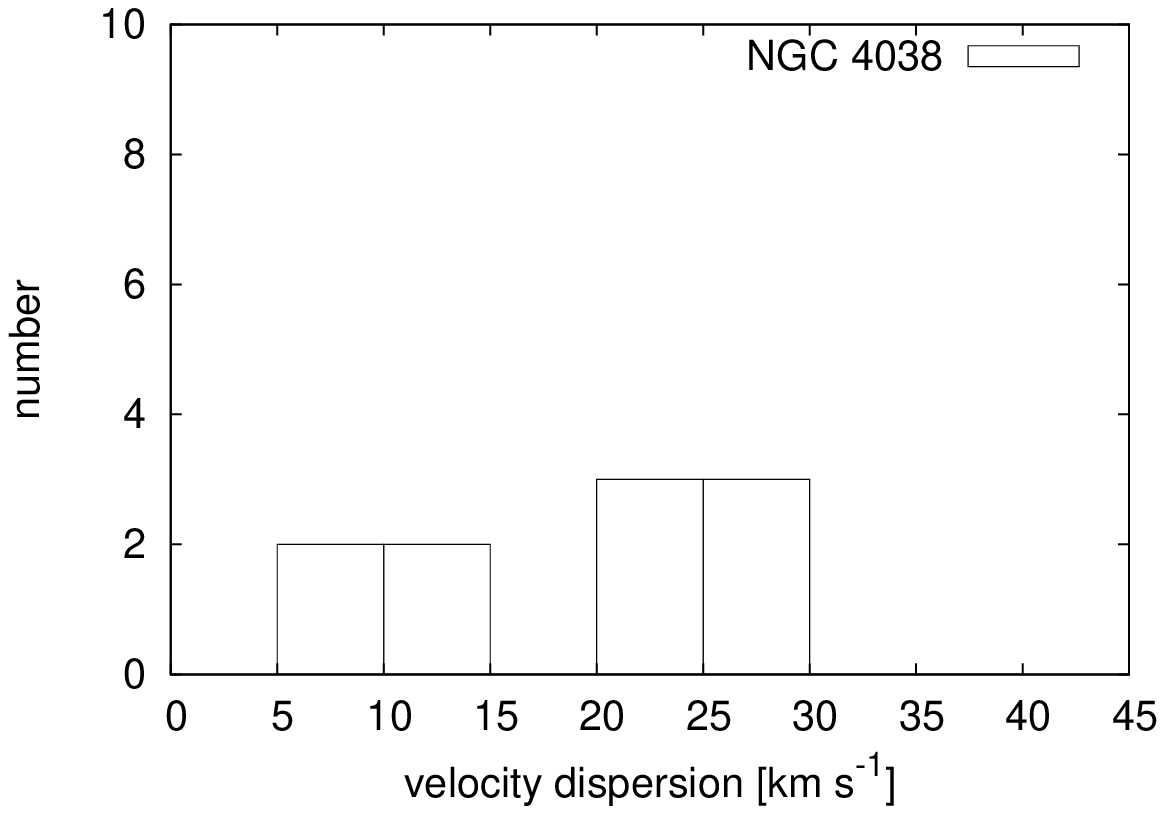}
  \end{center}
 \end{minipage}
 \begin{minipage}{0.3\hsize}
  \begin{center}
   \includegraphics[width=60mm]{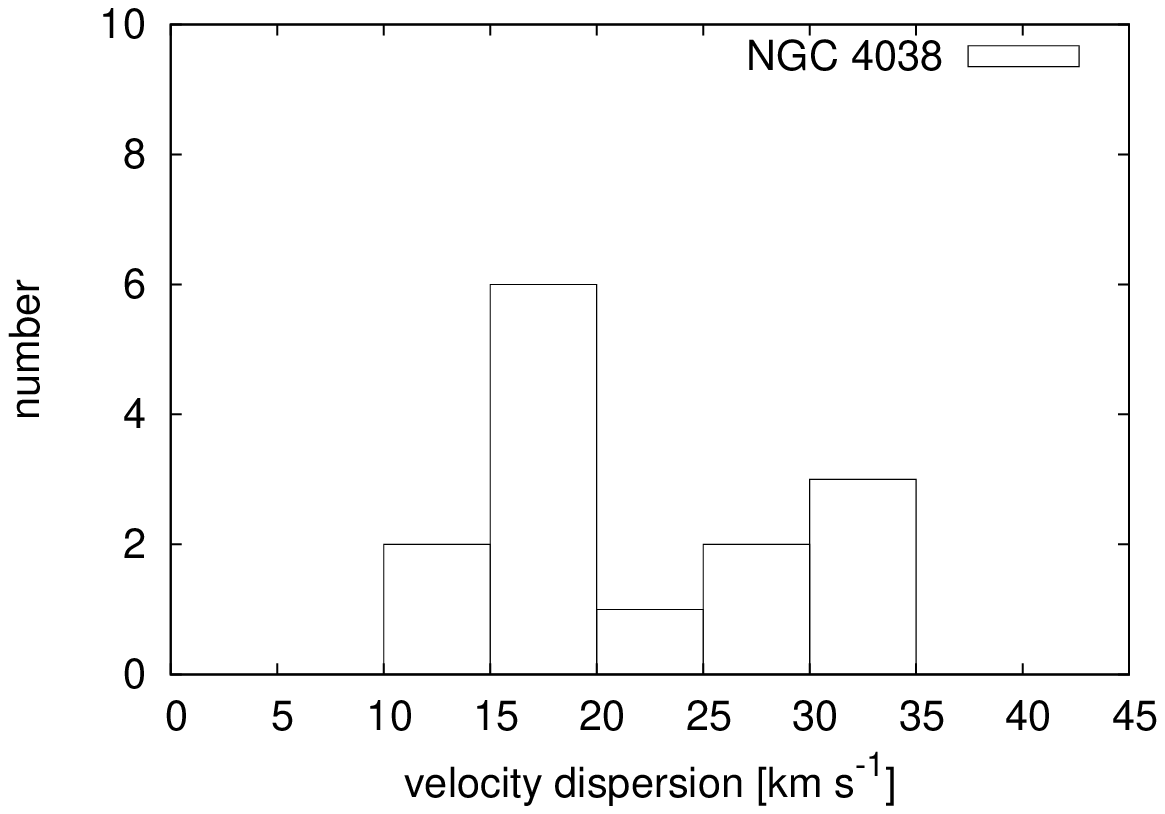}
  \end{center}
 \end{minipage}
 \begin{minipage}{0.3\hsize}
  \begin{center}
   \includegraphics[width=60mm]{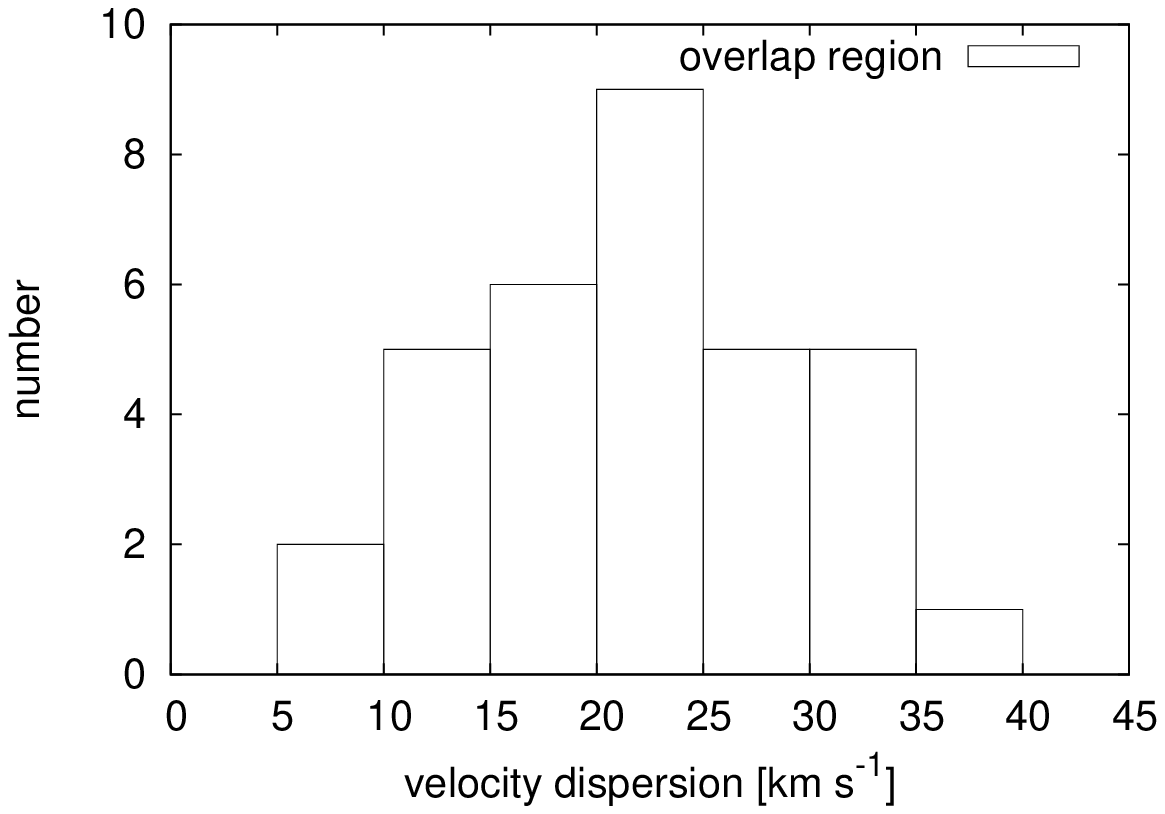}
  \end{center}
 \end{minipage}
 \caption{
	Distribution of the velocity dispersions of the detected molecular complexes 
	in NGC~4038 (left), NGC~4039 (middle) and overlap region (right).
 }
 \label{fig:sigma_hist}
\end{figure}

The velocity dispersions of the molecular complexes in each region 
are shown in Figure \ref{fig:sigma_hist}.  
We conducted a Kolmogorov-Smirnov test on the distribution of the velocity dispersions, 
and obtained P-value = 0.664 -- 0.973 (Table~\ref{tb:P-value}), 
suggesting that the velocity dispersions among the three different regions are similar.  
In NGC~4039, molecular complexes with larger velocity dispersions 
(25 -- 35 km s$^{-1}$) are located within 500 pc from the nucleus 
and those with a smaller velocity dispersion (10 -- 25 km s$^{-1}$) 
are located in the gas arm.  
Comparing the molecular complexes in the Antennae with clumps 
in M64 which is a molecule-rich spiral galaxy located at 4.1 Mpc \citep{Rosolowsky05} 
and the Galaxy \citep{Oka07}, the molecular complexes with the same velocity dispersion 
have an order of magnitude larger molecular mass in the Antennae.  

\subsubsection{The Relation between Radius and Velocity Dispersion} 
\begin{figure}[htbp]
\plotone{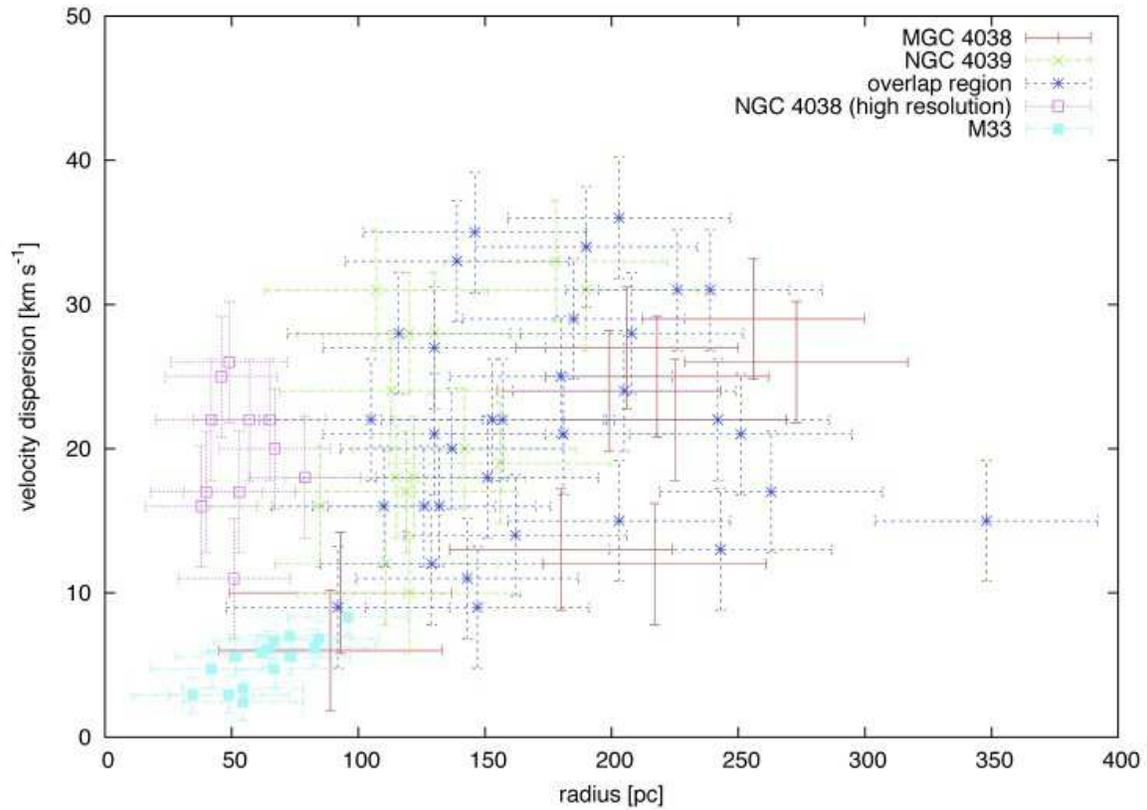}
 \caption{
	The relation between the radius and the velocity dispersion of the identified molecular complexes.  
	The data of clumps in M33 was obtained by \citet{Onodera09}.
 }
 \label{fig:r-sigma}
\end{figure}

On the GMC scales, it is known that 
the radius is proportional to the velocity dispersion \citep{Larson81}.  
\citet{Rosolowsky03} find a correlation between the radii and velocity dispersions 
of GMCs in M 33 which is a normal spiral galaxy located at 850 kpc, 
and the correlation appears to hold even in the outer disk \citep{Bigiel10}.  
In other instances, \citet{Fukui08} showed that the GMCs 
in the Large Magellanic Cloud (LMC) follow a similar correlation, 
but it is offset from that found for the GMCs in the inner Galaxy \citep{Solomon87}.  
In order to investigate if such correlation exists in the Antennae at GMA scales, 
we plot the relation between the radii and the velocity dispersions 
of the identified molecular complexes (Figure~\ref{fig:r-sigma}).  
The correlation coefficients derived from the least squares fit to the different regions 
are 0.68 in NGC~4038, 0.25 in NGC~4039, and 0.01 in the overlap region, 
suggesting the absence of a strong correlation in each region of the Antennae.  
At these scales, the velocity dispersion likely reflects the relative bulk motion 
of a collection of CO~(3--2) emitting clumps, and not the intrinsic velocity dispersions 
pertaining to the individual GMCs.  
Hence the absence of a strong correlation at size scales of GMAs is somewhat expected, 
especially in the dynamically disturbed parts of the galaxy such as the overlap region.  

We investigate further the relation between radius and velocity dispersion
of molecular complexes in NGC~4038 
because the correlation coefficient in NGC~4038 is the highest among three regions.  
By using Clumpfind on the high resolution data, 
we identified 11 molecular complexes that are mostly located 100~pc north of NGC~4038.  
Their radii range between 40 and  80~pc and are about the same size as the GMCs in M~33.  
The relatively tight correlation between radius and velocity dispersion is seen in the GMCs 
detected in a quiescent spiral galaxy M~33 \citep{Onodera09} (Figure~\ref{fig:r-sigma}).  
However, the velocity dispersions of the molecular complexes in NGC~4038 
are 2 --3 times larger than the GMCs in M~33 and distributed between 10 and 30 km s$^{-1}$.  
This is further evidence suggesting that the velocity dispersion does not reflect 
the intrinsic velocity dispersion but the relative bulk motion of CO~(3--2) emitting clumps.  

Finally, we note that the lack of significant difference 
in the velocity dispersion of molecular complexes 
between NGC~4038 and NGC~4039 may be a result 
of limited sensitivity and angular resolution.  
The clump identification routine is unable to separate the apparently smooth 
and continuous distribution of molecular gas into smaller clumps.

\subsubsection{Distribution of the Molecular Complexes and Their Relation to Star Clusters} 
\begin{figure}[htbp]
 \begin{center}
  \includegraphics[width=150mm]{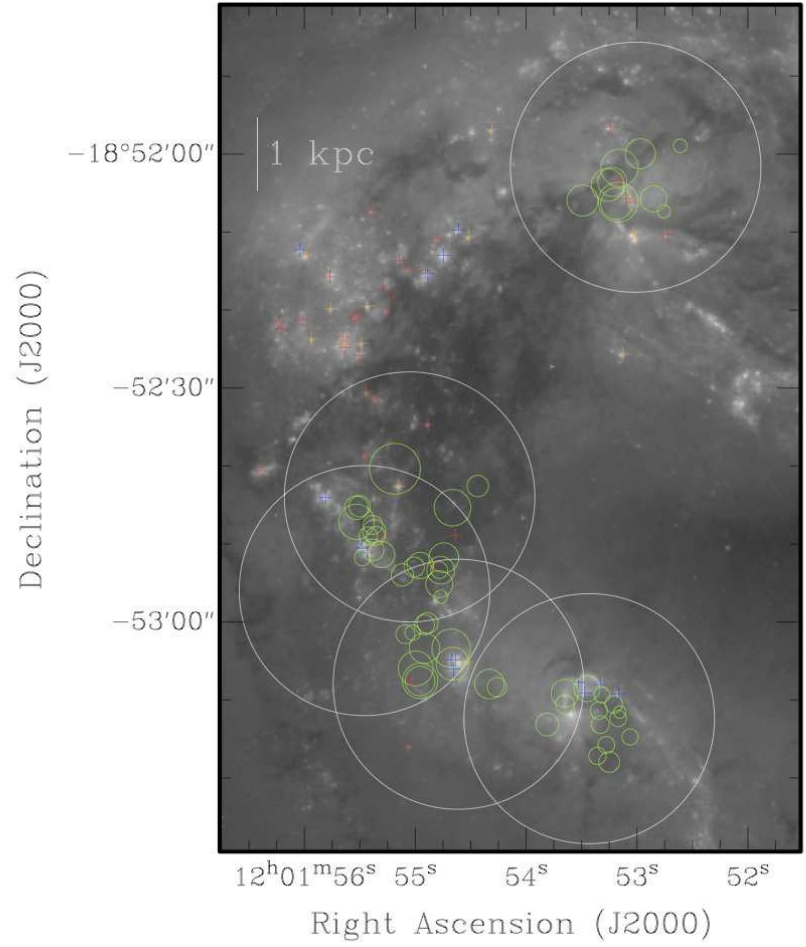}
 \end{center}
 \caption{
 	Distribution of the molecular complexes and their relation to star clusters.  
	The green circle signs show the position of the identified molecular complexes.  
	The circle sizes represent the size of the molecular complexes.  
	The blue, red and  orange cross signs show the location of star clusters classified 
	into the 50 most luminous (at V-band), most IR-bright clusters and both category, 
	respectively \citep{Whitmore10}.  The five white circle show the primary beam 
	centered on each pointing.  
 }
 \label{fig:cluster}
\end{figure}

The green circles in Figure \ref{fig:cluster} show the positions 
of the identified molecular complexes 
and the crosses represent the positions of 50 most luminous (at V-band) and/or 
the 50 most NIR-bright clusters whose ages are older than 1~Myr \citep{Whitmore10}.  
Overall, only 17 out of 57 molecular complexes (30 $\%$) host 
one or more of these star clusters with a median age of 4~Myr 
estimated from the ages of the star clusters \citep{Whitmore10}, 
not considering projection effects and extinction.  
In a related study, \citet{Mengel05} identified star clusters using the VLT K$_{s}$-band
and found that about 70 $\%$ of the star clusters 
with masses $\leq 10^{5}~M_{\sun}$ are younger than 10~Myr 
and most of the young clusters with the age $<$ 6~Myr are located in the overlap region.  
Although \citet{Mengel05} suggest that the CO~(1--0) obtained 
at a spatial resolution of $\sim 4 \arcsec$ in the overlap region 
shows spatial correlation with the location of the star clusters, 
our CO~(3--2) emission map with a spatial resolution of $\sim 1 \arcsec$ 
shows a clear offset from the star clusters, 
assuming that the star clusters identified using the V-band image 
overlap with those identified using the K$_{s}$-band image.  
The offset is evident in the southern part of the overlap region, 
where the CO~(3--2) emission is strong and the CO line ratio is relatively high.  
Hence, the absence of a significant spatial correlation 
between the locations of the molecular complexes and star clusters 
may suggest that the molecular gas traced in CO~(3--2) emission 
corresponds to the regions of future star formation, 
or regions where star formation occurred in the past $\sim$ 1~Myr 
because no star cluster which is younger than 1~Myr
correspond to the molecular complexes.  
This timescale is similar to the typical timescale for disassociation 
from the molecular cloud which is under 3~Myr \citep{Lada03}.

\subsection{Distribution of the integrated brightness temperature ratio} 
We investigated the integrated brightness temperature CO~(3--2)/(1--0) ratios 
using the SMA CO~(3--2) and OVRO CO~(1--0) data \citep{Wilson00}.  
While we assume the same beam filling factor for both transitions, 
it is possible that the true distribution of the CO~(3--2) emitting cloud is 
more clumpy and concentrated than the clouds emitting CO~(1--0) emission.  
Therefore, the peak line ratios we estimate here are lower limits.  
We clipped the visibilities so that both data have the same shortest uv range 
(minimum uv distance = 10.48 k$\lambda$ = 9.13~m) 
and then convolved the CO~(3--2) data to the angular resolution 
(5$\arcsec$.49 $\times$ 3$\arcsec$.84) of the CO~(1--0) image 
before estimating the line ratios.  
Finally, the line ratios are clipped at the 3$\sigma$ level.  

The resultant ratio map is shown in Figure~\ref{fig:ratio}.  
We separate the ratio map into five regions 
(NGC~4038, NGC~4039, Complex 1, Complex 2, and Complex 3) 
and estimate the mean ratio for each region.  
The mean integrated intensity ratio ranges from 0.3 to 0.6 (Table~\ref{tb:ratio}).  
Here the error bars are derived from the uncertainties in the flux calibration.  
On average, the CO line in the Antennae is not thermalized up to $J$ = 3--2.  
They are slightly less than the average ratios estimated 
by \citet{Zhu03} using single dish telescopes, 
which ranged between 0.67 and 0.91.  
In contrast, the line ratios are close to unity in the southern part of 
Complex 3 and the two nuclei.  
\citet{Iono09} derived 0.93 for the central regions of 12 dusty U/LIRGs 
in the local universe, which is also in agreement with 
the starburst galaxies Arp220 and M82 where the derived ratios 
are $\geq 1$ \citep[e.g.][]{Greve09}.  
Thus, the average excitation conditions analyzed at $\sim$ 1~kpc scale 
in the Antennae are close to the average properties of nearby galaxies 
\citep[e.g. 0.44 is the average ratio among 7 nearby galaxies;][]{Mao10}, 
except for two localized regions where the excitation conditions are similar 
to the central regions of starburst galaxies.  

Finally, we mention two issues that likely affect the analysis 
and interpretation of the line ratios.  
One issue is the inapplicability of the Rayleigh- Jeans approximation, 
especially for the higher frequency CO~(3--2) line.  
In such case the brightness temperature ratio can be less than unity 
(0.8 -- 0.9 for a temperature of 40 K; see \citet{Harris10}), 
assuming a thermalized and optically thick, single component cloud.  
The other possibility arises from the filling factor difference 
between the CO~(3--2) and CO~(1--0) lines. Such a multi-component cloud 
may be more realistic than adopting a single component cloud, 
and higher angular resolution ALMA/ACA observations of both lines 
should provide us with the exact distribution of each line tracers in the future.

\subsubsection{The Nucleus of NGC~4038/9} 
The mean $^{12}$CO~(3--2)/(1--0) ratio of NGC~4038 is 0.6 $\pm$ 0.2, 
which is the highest among the averages of the five complexes.  
The spatial distribution of the ratio presented in Figure~\ref{fig:ratio} 
shows that the ratio is 0.75 in the center, 
and this is two times higher than the ratios in the outer region.  
The regions with particularly high line ratios correspond well 
with the peaks of the MIR and NIR emission, 
qualitatively suggesting the correlation 
between high line ratio and embedded star formation.  
This is consistent with observations in nearby galaxies 
where the ratio appears to increase from the outer disk to the central regions
 \citep[e.g.][]{Israel09, Wilson09, Boone11}.  
A similar trend is also seen in the Milky Way galaxy, 
where the ratio increases from $\sim 0.5$ in the inner disk to $\sim 0.9$ 
towards the galactic center where the star formation is more active \citep{Oka07}.  
Furthermore, signatures of more recent star formation activity 
in NGC~4038 are seen in optical and X-ray observations 
\citep[e.g.][]{Weedman05, Zezas06, Klaas10}, 
suggesting the coexistence of 
a variety of star clusters formed at various stages of the tidal activity.  

The line ratio near the nuclear region of NGC~4039 is high ($\sim 1$), 
despite a significantly lower mean ratio (0.5$\pm$0.1).  
The star formation activity in the central region of NGC~4039 is 
only modest \citep[SFR = 0.50 M$_{\odot}$ yr$^{-1}$;][]{Klaas10}, 
which is only half of the value derived in NGC~4038.  
The absence of the Br$_{\gamma}$ line in the stellar continuum 
is an indication of the dominance of old giants and red supergiants, 
further suggesting inactivity \citep{Gilbert00}.  
Thus it is not likely that the high line ratio in the nuclear region is 
produced by current star formation.  

The CO line could be collisionally heated and excited up to high energy levels 
if NGC~4039 harbors an AGN \citep{Matsushita04, Krips11}.  
A bright diffuse X-ray emission region is associated with the K$_{s}$-band nucleus 
within 220 pc, and also with the region with high line ratio.  
The radio continuum has a steep spectrum, 
which suggests non-thermal emission arising from supernova remnants 
\citep{Neff00}, or X-ray binaries \citep{Zezas02a, Zezas02b}.  
Furthermore, \citet{Brandl09} found that the observed infrared continuum 
is consistent with a pure starburst model and 
concluded that the spectrum of the nucleus is produced 
by emission from dust that is heated by star formation alone.  
Thus, these multi-wavelength properties of NGC~4039 provide evidence 
against the presence of an AGN.  

Another possible cause for the high line ratio is gas collision arising 
from violent gas kinematics.  
As shown in \S 3, the distribution of the CO~(3--2) gas in NGC~4039 
is composed of two main components; the nuclear concentration and 
the single-arm gas that appears to stream along the optical dust lane.  
Although the spatial resolution of the ratio image is limited to 500 pc, 
the region where the two components connect is consistent 
with the position that displays the highest ratio.  
The properties of the molecular complexes located at this region are 
markedly different from the average properties of NGC~4039, 
where the radii of the two molecular complexes are larger.   
In addition, the velocity dispersion of the molecular complexes 
in this regions are $\sim 50\%$ larger than the average of NGC~4039.  
Furthermore, the recession velocities (see Table \ref{tb:clump})
of these molecular complexes are consistent 
with the velocity of the clouds with the highest ratio.  
This observational evidence suggests a scenario 
where the molecular gas flows into the central disk (see \S4.4), 
collides, and results in high excitation.  
Future high resolution observations of shock tracers 
is necessary to confirm this scenario.

\subsubsection{The Overlap Region} 
\begin{figure}[htbp]
\plotone{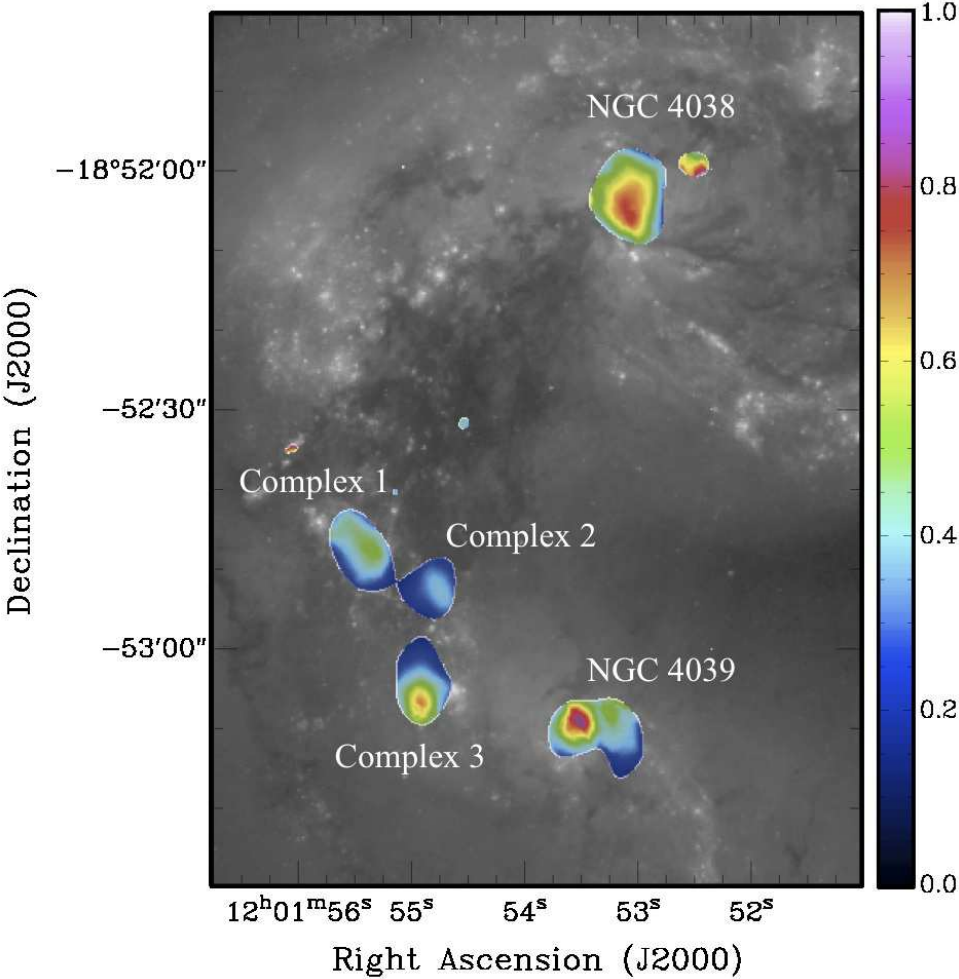}
 \caption{
	The integrated brightness temperature CO~(3--2)/(1--0) ratios.  
	The background image is the HST 435 nm \citep{Whitmore99}.  
 }
 \label{fig:ratio}
\end{figure}

The mean $^{12}$CO~(3--2)/(1--0) ratio is 0.4$\pm$ 0.1 in Complex 1, 
0.3$\pm$ 0.1 in Complex 2, and 0.5$\pm$ 0.1 in Complex 3.  
Although the average ratios are all consistent within the errors, 
gradients in the line ratio are seen in each complex (Figure~\ref{fig:ratio}), 
with a particularly steep gradient in Complex~3.  
\citet{Zhang10} derived the FUV -- 24 $\mu$m broad-band spectral energy distribution 
(SED) and found that the UV/optical emission gradually become redder and weaker 
from Complex~1 (which corresponds to their region~6) to Complex~3 (region~4).  
This suggests that the ratio of young star population ($<$ 10~Myr) 
to intermediate/old star population ($>$ 10~Myr) increases from Complex~1 to 3.  
The continuum emission at MIR wavelengths in Complex~1 (SGMC~1 in \citet{Wilson00}) 
and Complex~2 (SGMC~2) are flat, but Complex~3 (SGMC 3--5) contains strong 
and rising continuum emission \citep{Wilson00}.  
The mid-infrared continuum emission in Complex~3 can be produced by massive O stars.  
Similarly, strong FIR continuum emission seen in Complex~3 (K1) has 
the highest SFR and SFE and maximum dust temperature \citep{Klaas10}.  
This observational evidence suggests that Complex~3 is likely to be 
a very young complex and star forming histories are different from Complex~1 and 2.  
This is consistent with the higher CO line ratio in Complex~3.  

\begin{figure}[htbp]
 \begin{center}%
  \includegraphics[width=80mm]{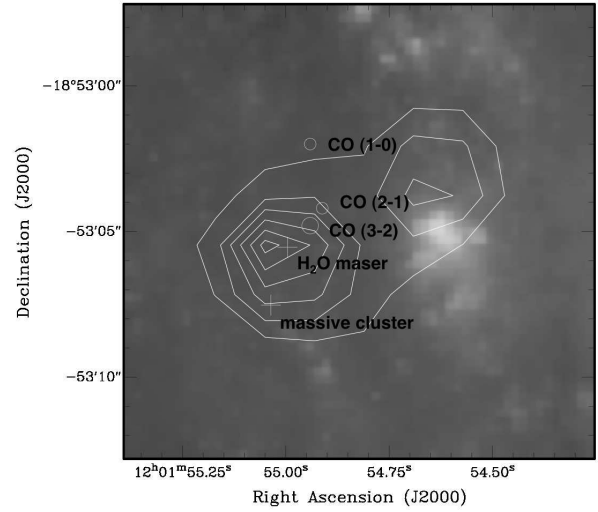}
 \end{center}
 \caption{
	The closeup of the Complex 3.  
	The Spitzer~8 $\mu$m contour map overlaid on the HST H$\alpha$ image.  
	The contour levels are 40, 50, 60, 70, 80, and 90 $\%$ of the peak intensity.  
	The upper, middle, and lower circles show the CO~(1--0), the CO~(2--1), and the CO~(3--2) peaks, respectively.  
	The upper cross shows the location of the H$_{2}$O maser observed by \citet{Brogan10} and 
	the lower cross shows the location of the most massive cluster identified by \citet{Whitmore10}.  
 }
 \label{fig:Complex3}
\end{figure}

The overlap region of the Antennae was observed with a single pointing 
in the $^{12}$CO~(2--1) line at the SMA.  
From our analysis of the archival data, the distribution 
of the CO~(3--2) and CO~(2--1) maps are similar overall, 
but subtle differences are seen particularly in Complex~3.  
We found that the CO~(2--1) peak in Complex~3 is located 
100~pc north of the CO~(3--2) peak, 
and the CO~(1--0) peak is further displaced toward the north 
by 200~pc as shown in Figure \ref{fig:Complex3}.  
\citet{Brogan10} detected an H$_{2}$O maser within 100~pc 
of the CO~(3--2) peak, and estimated from the VLA 3.6~cm continuum 
that the ionized gas is equivalent to 2000 -- 5000 O$^{*}$ stars.  
Furthermore, the most massive cluster is embedded 350~pc southeast 
of Complex 3 \citep{Whitmore10}.  
These are observational indicators that suggest the earliest stages 
of active star formation in the southern part of Complex~3 
are entangled with more mature clusters that were formed 1~Myr ago.  
The high line ratio coincides with this region, suggesting that
heating of the interstellar medium due to UV emission from newly born stars 
is responsible for collisionally exciting the CO gas to the $J$=3--2 level.  

\subsection{Evidence for a Counter Rotating Nuclear Disk and Gas Inflow Toward NGC~4039} 
\begin{figure}[htbp]
 \begin{minipage}{0.5\hsize}
  \begin{center}
   \includegraphics[width=80mm]{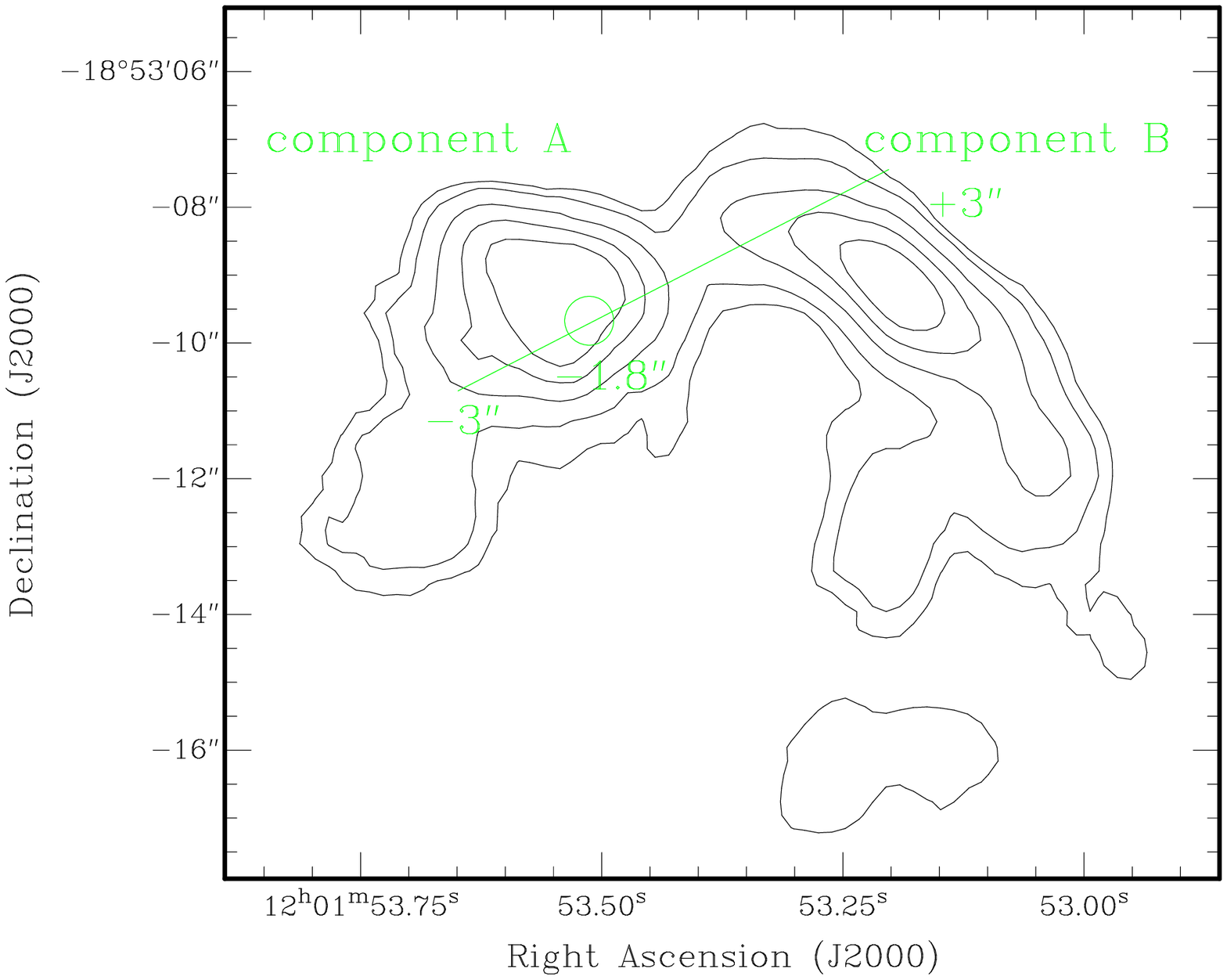}
  \end{center}
 \end{minipage}
 \begin{minipage}{0.5\hsize}
  \begin{center}
   \includegraphics[width=80mm]{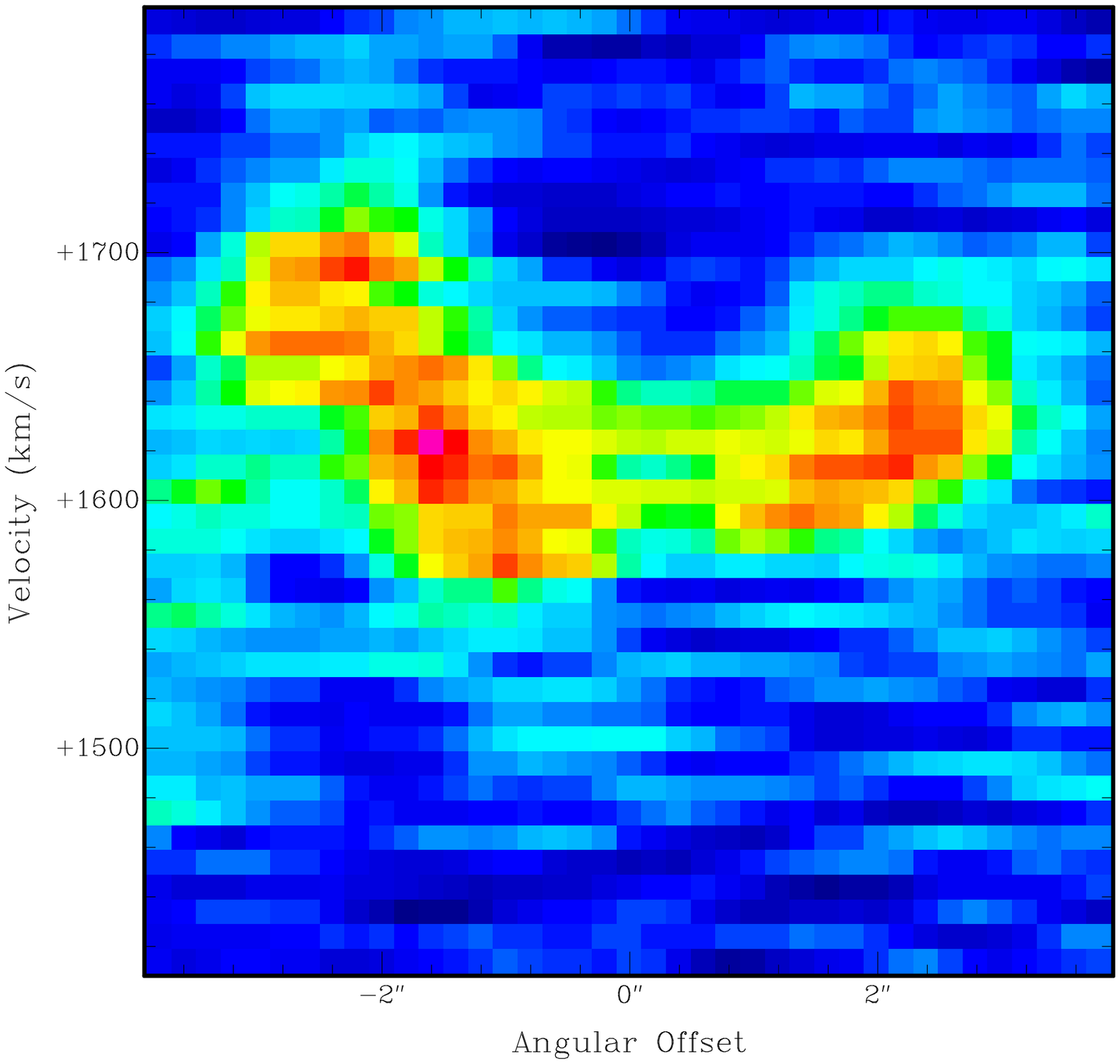}
  \end{center}
 \end{minipage}
 \caption{
	The right figure is the PV diagram along the green cut line 
	shown in the left figure.  The position whose corresponds 
	to the VLT K$_{s}$-band peak of NGC~4039 shown the green circle 
	in the left figure.  The contour levels are 6 Jy km a$^{-1}$ $\times$ 2, 3, 5, 7, 9.  }
 \label{fig:pv_n9}
\end{figure}

The spatial distribution of the CO~(3--2) emission in NGC~4039 
(Figure \ref{fig:pv_n9} (left)) clearly shows two distinct molecular gas components 
that are separated by a projected distance of 560~pc.  
One component (component A; see Figure \ref{fig:pv_n9}(left)) is 
apparently distributed around the nucleus and the other component 
(component B; see \ref{fig:pv_n9}(left)) is 
distributed along the dust lane seen in the B-band image.  
The masses of components~A and B are 5.2 $\times$ 10$^{8}$ 
and 4.5 $\times$ 10$^{8}$ M$_{\sun}$, respectively.  
The PV diagram of component A is shown in Figure \ref{fig:pv_n9} (right).  
The systemic recession velocity of NGC 4039 determined 
from our data is 1610 km s$^{-1}$.  
The mean gas velocity of component A is blue- and redshifted 
by -50 and +30 km s$^{-1}$ with respect to the K$_{s}$-band nucleus 
to the west and east side, respectively.  
Then assuming that component~A is in the plane of the galaxy, 
this suggests that component~A rotates around the galactic center of NGC~4039.  
We compare the dynamical mass and molecular mass of component~A 
and find that the disk is dynamically stable, 
assuming that component~A is a disk with a rotation velocity of 40 km s$^{-1}$ 
and the disk inclination of $i$ = 60$\arcdeg$ \citep{Toomre72}.  
\citet{Downes98} found a high occurrence of rotating 100~pc scale nuclear disks 
or rings from their molecular gas observations of ULIRGs.  
While the kinematic signature of a disk in NGC~4039 is less obvious 
than the ULIRGs in \citet{Downes98} mainly because of a lower S/N ratio, 
presence of such nuclear disks may be a common phenomenon in the nuclear 
regions of colliding systems.  
A clear characteristic seen in NGC~4039 is that 
this component appears to rotate in the opposite direction 
from the galactic rotation defined by the direction 
of the spiral arms seen in optical images, 
which is defined from the direction of the trailing arm.  

\begin{figure}[htbp]
 \begin{minipage}{0.5\hsize}
  \begin{center}
   \includegraphics[width=80mm,clip,trim=0 0 0 0]{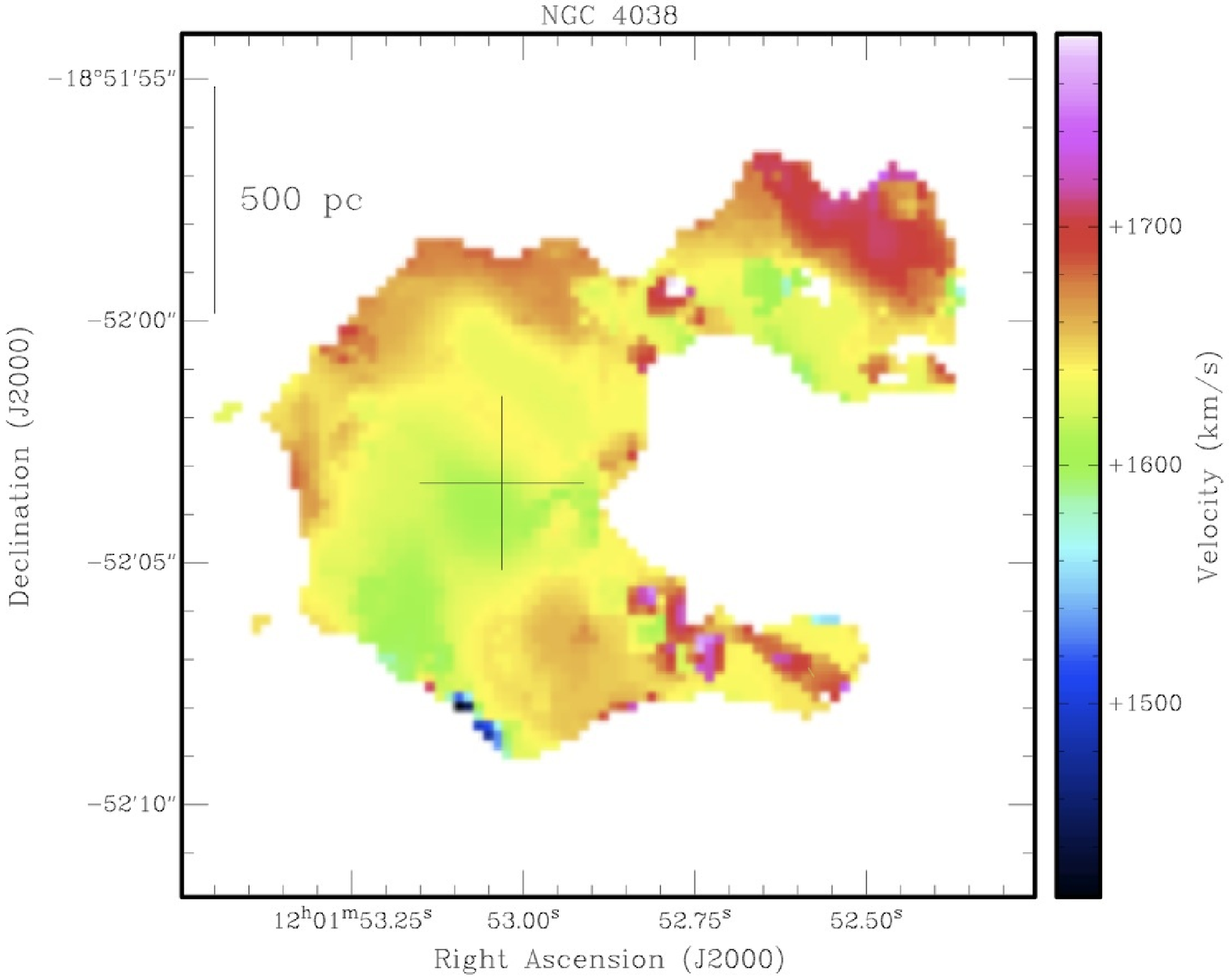}
  \end{center}
 \end{minipage}
 \begin{minipage}{0.5\hsize}
  \begin{center}
   \includegraphics[width=80mm,clip,trim=0 0 0 0]{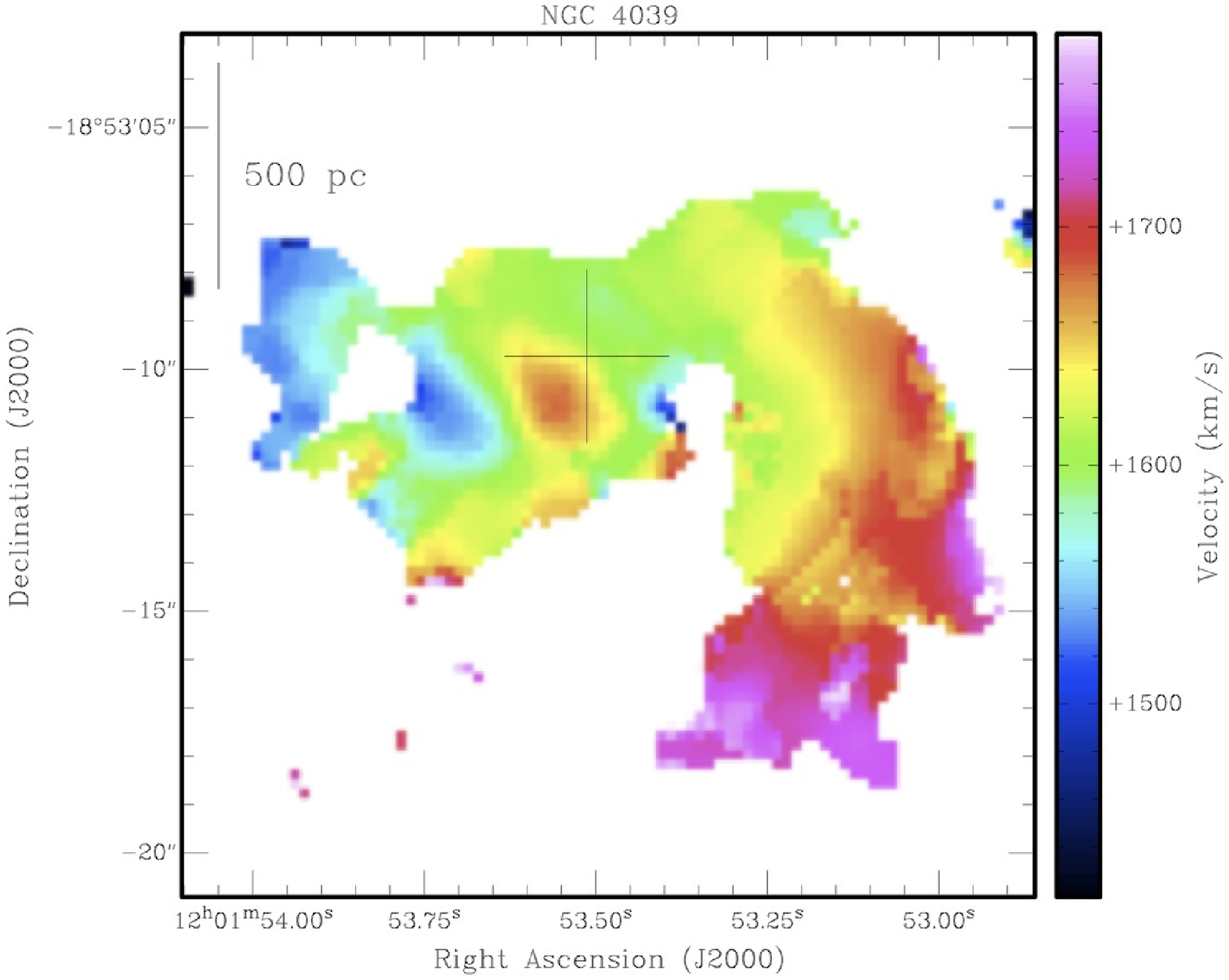}
  \end{center}
 \end{minipage}
 \begin{minipage}{0.5\hsize}
 \begin{center}
  \includegraphics[width=85mm,clip,trim=15 0 0 0]{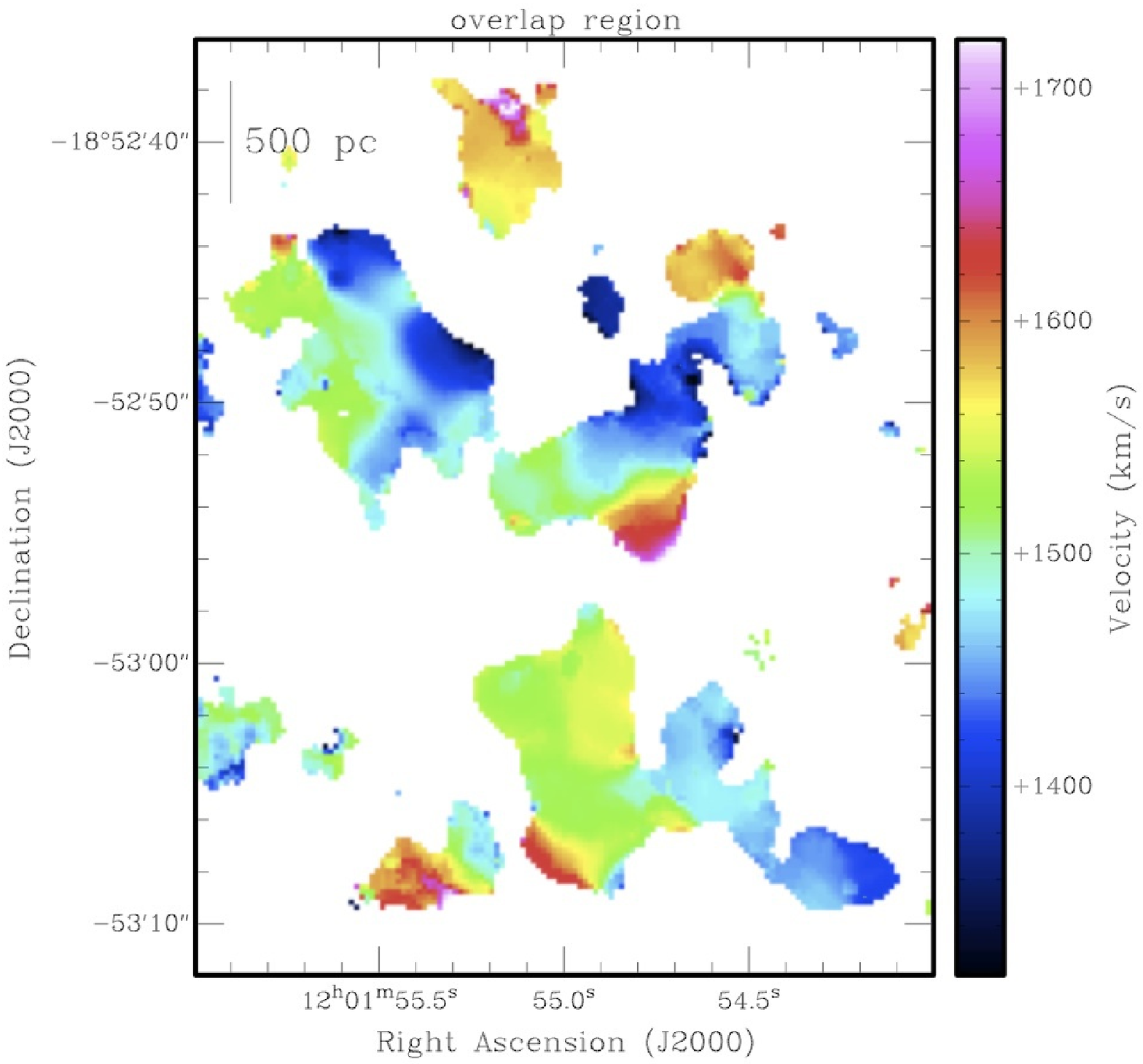}
 \end{center}
 \end{minipage}
 \caption{
	The velocity field of NGC~4038 (upper-left), NGC~4039 (upper-right), 
	and the overlap region (bottom).  The cross signs in the two upper 
	figures show the galactic center defined by the VLT K$_{s}$-band image.  
 }
 \label{fig:mom1}
\end{figure}

$N$-body/smoothed particle hydrodynamic (SPH) simulation performed 
by \citet{Barnes96} predicts that a non-axisymetric potential can enhance 
the stellar torques exerted on the gas, 
causing gas inflows into the central regions in a 
merging process. These inflows may be kinematically distinct, 
as shown in the simulation studies by \citet{Iono04}.  
Component B shows the predicted kinematic signature 
of gas streaming along the arm ($\sim$ 1 kpc), 
as shown in  Figure \ref{fig:mom1}.  
In addition, a high line ratio is seen at the location 
where Components~A and B connect 
and the velocity dispersion of the molecular complexes 
are $\sim 50\%$ larger than the average of NGC~4039
as shown in  \S 4.2.1.  
If Component~B is indeed related to gas inflow, 
it is then possible that high gas excitation is caused 
by collisions between the inflowing gas component and the counter-rotating disk.  

\subsection{A possible molecular gas bubble in NGC~4038} 
\begin{figure}[htbp]
 \begin{center}
  \includegraphics[width=100mm]{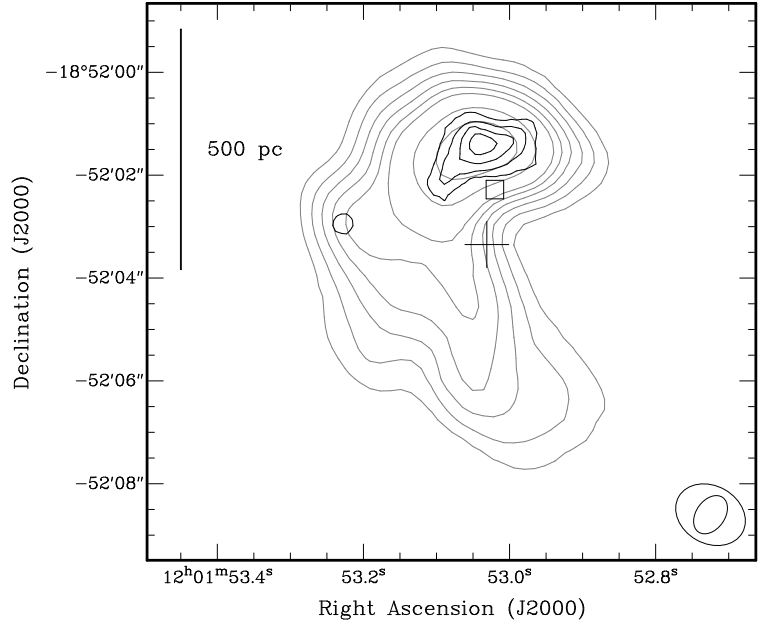}
 \end{center}
 \caption{
	CO (3-2) integrated intensity contour map in NGC 4038.  The black contours show 
	the CO~(3--2) emission using only the data observed in the extended configuration.  
	The contour levels are 5~Jy km s$^{-1}$ $\times$ 5, 7, 9, 11.  The small ellipse 
	in the lower right corner indicates the higher angular resolution ($0\farcs82 \times 0\farcs65$).  
	The grey contour map is the CO~(3--2) map made with all the data.  
	The contour levels are 6~Jy km s$^{-1}$ $\times$ 5, 7, 9, 11, 13, 15, 20, 25.  
	The large ellipse in the lower right corner indicates the angular resolution 
	($1.42\arcsec \times 1.12\arcsec$).  
	The cross sign shows the galactic center defined by the VLT K$_{s}$-band image 
	and the square sign shows the nonthermal radio source \citep{Neff00}.  
} 
 \label{fig:extended}
\end{figure}

The high resolution image ($\sim 0\arcsec7$; Figure \ref{fig:extended}) 
allows us a detailed investigation of the molecular gas 
surrounding the nuclear region of NGC~4038.  
The molecular and virial masses of the central component 
are 3.2 $\times$ 10$^{8}$ M$_{\sun}$ and 2.0 $\times$ 10$^{8}$ M$_{\sun}$, 
respectively, suggesting that this component is gravitationally bound.  
Furthermore, by using Clumpfind on the high resolution data, 
we identified 11 molecular complexes that are mostly located 
100~pc north of the galaxy center.  
The virial ratios for all of the clumps are less than unity, 
suggesting gravitationally bound clouds.  

A nonthermal radio source with an energy equivalent to
115 times Cassiopeia A is located 110~pc north of the K$_{s}$-band peak
(square sign in Figure~\ref{fig:extended}) (Neff $\&$ Ulvestad 2000).  
The distribution of CO~(3--2) appears to form an arc 
centered around this radio source, with an apparent absence 
of significant CO~(3--2) emission at the radio source.
Similar molecular bubbles have been found in M~82 \citep[e.g.][]{Matsushita04} 
and NGC~253 \citep{Sakamoto06}.  
They attributed it to supernova blasts
that swept away the molecular gas from the surrounding medium.  
Following \citet{Sakamoto06}, we estimated the kinetic 
energy needed to create the bubble using, 
$E \approx (10\pi/3) \rho v^{2} R^{3}\hspace{0.2cm}$, 
where $\rho = 1.4 n_{0} m_{H}$ ($n_{0}$ is the hydrogen number density
of the prebubble medium, and $m_{H}$ is the the atomic mass of hydrogen) 
is the initial mass density of the interstellar matter, $R$ is the size of a bubble, 
and $v$ is the expansion velocity of molecular gas.  
We derived $R$ from the distance between the nonthermal radio source 
and the CO (3--2) peak in Figure \ref{fig:extended}, which is 110~pc, 
and used the offset from systemic velocity at the CO (3--2) peak as $v$ 
of 30 km~s$^{-1}$.  
The estimated energy is $9 \times 10^{46}$~J, 
adopting $n_{0} = 10^{2}$~cm$^{-3}$, which is a typical density 
of interstellar matter in the Galaxy and nearby galaxies.  
This energy corresponds to the kinetic energy released 
by 10 -- 100 supernova explosions 
and also to the released energy estimated from the radio continuum.  
Thus a high frequency of recent supernova explosions is a possible cause 
of the apparent absence in the CO~(3--2) distribution.

\section{Summary} 
We present interferometric $^{12}$CO~($J$ = 3--2) observations 
toward main bodies of the Antennae galaxies (NGC~4038/9) 
obtained using the SMA.  
The image shows gas concentration at both nuclei and the overlap region, 
and the emission is roughly distributed along the dust lanes 
seen in the B-band image.  
The integrated intensity over main body of the Antennae 
is 4.4 $\times$ 10$^{3}$ Jy km s$^{-1}$ 
and the overlap region contains almost half of the emission.  
We detect using the Clumpfind algorithm 57 molecular complexes, 
33 of which located in the overlap region.  
Adopting a galactic CO-to-H$_{2}$ conversion factor 
and the estimated CO~(3--2)/(1--0) line intensity ratios, 
the molecular complexes appear to have masses of 10$^{7}$ -- 10$^{8}$ M$_{\sun}$.  
While we find that 58 $\%$ (33/57) of the spatially resolved GMAs 
are located in the overlap region, only $\leq 30 \%$ are spatially coincident 
with the star clusters detected in the optical and/or NIR images, 
suggesting that the bulk of the CO~(3--2) emission traces the regions 
with very recent or near future star formation activity.  

We find that the GMAs in the Antennae do not follow 
the well known radius-velocity dispersion relation 
obeyed by GMCs in Milky Way and other nearby galaxies.  
The molecular complexes with large radii are distributed 
in NGC~4038 and the southern part of the overlap region,
where the star formation rates are high.  
Therefore, the larger molecular complexes, 
which are possibly formed by clustering around local peaks 
of the gravitational potential wells, 
may be used as the signposts of current dusty star forming regions.  
In NGC~4039, the molecular complexes with the larger velocity dispersion 
(25--35 km s$^{-1}$) are located within 500 pc of the nucleus 
and those with the smaller velocity dispersion (10--25 km s$^{-1}$) 
are located in the gas arm.  

We estimated the CO~(3--2)/(1--0) brightness temperature ratio 
using the SMA CO~(3--2) and OVRO CO~(1--0) data, 
assuming a same beam filling factor for both transitions.  
The mean ratios range between 0.3 and 0.6, which suggests that, 
on average, the CO line in the Antennae is not thermalized up to $J$ = 3--2.  
However the line ratios in two nuclei and the southern of the overlap region 
are close to unity in some localized regions.  
The high ratio in the nucleus of NGC~4038 and the overlap region 
is likely caused by star formation because the line ratio becomes high 
for clouds that are close to active star forming regions.  
In contrast, the nucleus of NGC~4039 has a relatively inactive star forming region.  
One possibility for causing the high line ratio in the nucleus of NGC~4039 
is a presence of a hidden AGN.  
The other is gas collision between the inflowing gas 
and the counter-rotating central disk.  

The Atacama Large Millimeter/Submillimeter Array (ALMA) 
will begin scientific observations in the second half of 2011.  
The Antennae is one of the best targets observed using the ALMA 
because it passes near zenith at the ALMA site 
and the beam pattern will be a nearly circular shape.  
Using the superb sensitivity and angular resolutions offered by ALMA, 
we can observe other molecular lines including high-$J$ CO 
in the Antennae and study the properties of molecular gas quantitatively. 

\acknowledgments
We thank the anonymous referee for useful comments.
We also thank Sachiko Onodera for providing the data of GMCs in M~33 
and giving helpful comments, which improved our discussion.

\clearpage

\begin{table} [htbp]
\begin{center} 
\caption{The phase centers}
\label{tb:fov}  
\begin{tabular}{ccc}
   \hline\hline
    & RA (J2000) & Dec (J2000) \\
   \hline
   FOV 1 & 12:01:53.45 & -18:53:12.40 \\
   FOV 2 & 12:01:54.61 & -18:53:08.00 \\
   FOV 3 & 12:01:55.46 & -18:52:55.98 \\
   FOV 4 & 12:01:55.05 & -18:52:43.96 \\
   FOV 5 & 12:01:53.01 & -18:52:01.68 \\
   \hline   
\end{tabular}
\end{center}
\end{table}

\begin{table}[htbp]
\begin{center} 
\caption{KS-test P-value}
\label{tb:P-value}
\begin{tabular}{ccc}
   \hline\hline
   regions & radius & velocity dispersion\\
   \hline
   (NGC~4038, NGC~4039) & 0.002 & 0.771 \\
   (NGC~4038, overlap region) & 0.199 & 0.664 \\ 
   (NGC~4039, overlap region) & 0.006 & 0.973 \\
   (NGC~4038, the southern part of the overlap region) & 0.975 & 0.675 \\
   (NGC~4039, the southern part of the overlap region) & 0.017 & 0.997 \\
   \hline
\end{tabular}
\end{center} 
\end{table}

\begin{table}[htbp]
\begin{center}
\caption{The mean integrated intensity ratio of each region  } 
\label{tb:ratio}
\begin{tabular}{cccccc}
   \hline\hline
    & mean ratio \\
   \hline
   NGC~4038 & 0.6 $\pm$ 0.2\\
   NGC~4039 & 0.5 $\pm$ 0.1\\
   Complex 1 & 0.4$\pm$ 0.1\\
   Complex 2 & 0.3$\pm$ 0.1\\
   Complex 3 & 0.5$\pm$ 0.1\\
   \hline
\end{tabular}
\end{center}
\end{table}

\clearpage
\makeatletter
    \renewcommand{\thetable}{%
    \thesection.\arabic{table}}
   \@addtoreset{table}{section}
  \makeatother
\makeatletter
    \renewcommand{\thefigure}{%
    \thesection.\arabic{figure}}
    \@addtoreset{figure}{section}
  \makeatother
  
\appendix
\section{Multiwavelength Images}
We compare the CO~(3--2) map with multiwavelength images from radio and to X-ray.  
Figure \ref{fig:stamp} shows the CO~(3--2) contour maps of five regions (NGC~4038, 
NGC~4039, Complex 1, Complex 2, and Complex 3) overlaid on 14 different images.  

The CO~(1--0) map was taken with the OVRO interferometer by \citet{Wilson00}.  
The strongest emission in all regions is located in NGC~4038 
and the emission in Complex~2 is strongest in the overlap region.  
\citet{Klaas10} conducted FIR observations at 70, 100 and 160 $\mu$m 
using the \textit{Herschel}-PACS.  
The brightest emission in three bands comes from Complex~3.  
Complex~1 and Complex~2 together become the brightest area 
in the 160 $\mu$m map.  
The 24 $\mu$m and four near-infrared (NIR) at 3.6, 4.5, 5.8, and 8.0 $\mu$m images 
were obtained with the MIPS \citep{Rieke04} and the IRAC \citep{Fazio04} 
on board the \textit{Spitzer Space Telescope}, respectively.  
We use the basic calibrated data obtained using the LEOPARD software.  
The strongest emission in all bands, especially 24 $\mu$m, is located in Complex~3.  
The H$\alpha$ image (F656N), and the B-band image (F435W) have been observed 
using the ACS on board the HST \citep{Whitmore10}.  
We use the archival data obtained from the MAST (Multimission Archive at STScI).  
There is no optically bright emission in Complex~2 and Complex~3.  
Both the far-ultraviolet (FUV; $\sim$ 1516 $\AA$) and the near-ultraviolet (NUV; $\sim$ 2267 $\AA$ ) 
images were taken with the \textit{Galaxy Evolution Explorer}~(GALEX) Ultraviolet Atlas of Nearby 
Galaxies distributed by \citet{Gil07}.  
The strong UV emission is detected west of the peak in Complex~3, 
where the bright optical, NIR and MIR emission is also observed.  
We obtained the archival X-ray image which \citet{Fabbiano03} observed with \textit{Chandra}.  
The bright X-ray emission is associated with the nucleus of NGC~4039.  

\begin{figure}[htbp]
 \begin{center}
  \includegraphics[scale=1.2,angle=270,clip,trim=0 0 0 379.5]{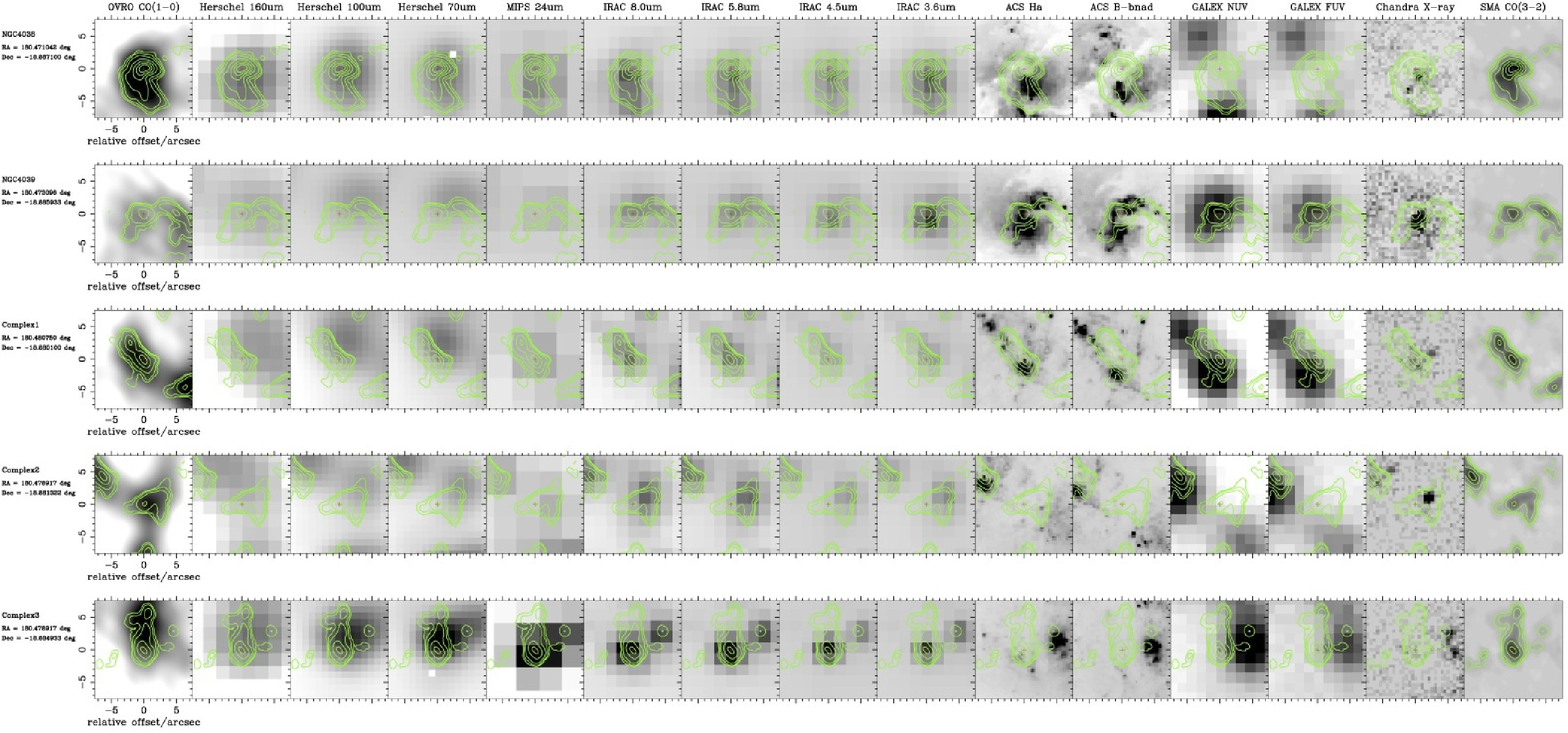}
 \end{center}
 \caption{
   The back ground images are the CO~(1--0), the 160$\mu$m, the 100$\mu$m, 
   the 70$\mu$m, the 24$\mu$m, the 8.0$\mu$m, and the 5.6$\mu$m images from left to right.  
   The contour maps are CO~(3--2) emission and the contour levels are 6 Jy $\times$ 2, 3, 5, 10, 15, 20, 25.  
   The red cross signs show the CO~(3--2) peaks each region.  The length of each side corresponds to 1.6 kpc.  
   Note that the coordinates of the 160$\mu$m, the 100$\mu$m, and the 70 $\mu$m maps 
   have a few arcseconds deviations from the coordinates of the CO~(3--2) map.  
 }
 \label{fig:stamp}
\end{figure}

\newpage
\addtocounter{figure}{-1}
\begin{figure}[htbp]
 \begin{center}
   \includegraphics[scale=1.2,angle=270,clip,trim=0 0 0 379.5]{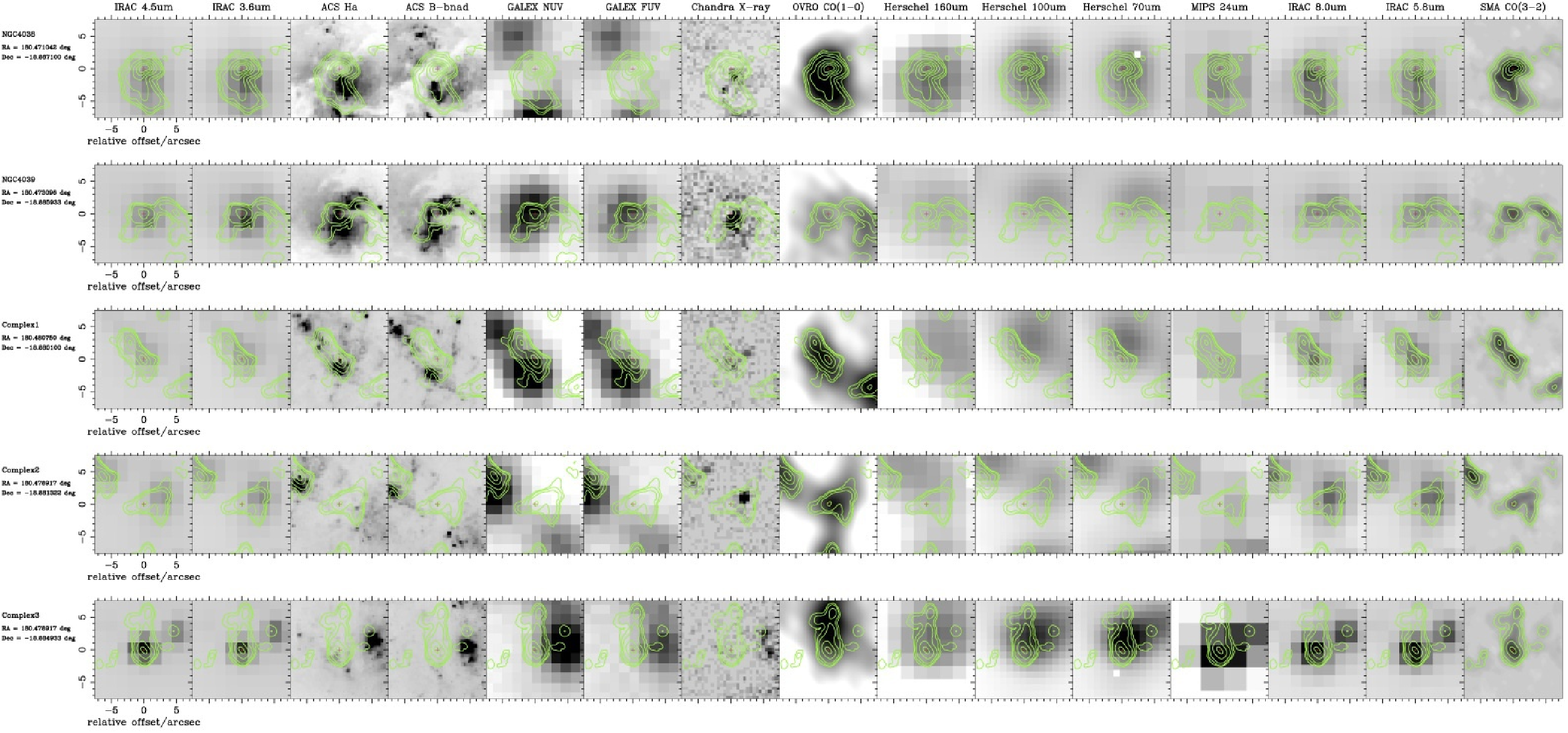}
 \end{center}
 \caption{
   Continued.  The background images are the 4.5$\mu$m, the 3.8$\mu$m, the H$\alpha$, 
   the B-band, the NUV, the FUV, and the X-ray images from left to right.  
 }
\end{figure}

\clearpage
\section{Catalog of the Molecular Complexes}
\begin{center}
\begin{longtable}{cccccccc}
\label{tb:clump} \\

\multicolumn{8}{c}
{{\bfseries \tablename\ \thetable{} -- The molecular complexes}} \\
\hline\hline
ID & RA (J2000) & Dec (J2000) & $V$ & $r$ & $\sigma_{v}$ & $M_{co}$ & $M_{vir}$ \\
 & [h m s] & [$^{\circ}$ $^{\prime}$ $^{\prime\prime}$] & [km s$^{-1}$] & [pc] & [km s$^{-1}$] & [$10^{7}$ M$_{\odot}$] &  [$10^{7}$ M$_{\odot}$] \\
\hline
\endfirsthead

\multicolumn{8}{c}
{{\bfseries \tablename\ \thetable{} -- Continued}} \\
\hline\hline
ID & RA (J2000) & Dec (J2000) & $V$ & $r$ & $\sigma_{v}$ & $M_{co}$ & $M_{vir}$ \\
 & [h m s] & [$^{\circ}$ $^{\prime}$ $^{\prime\prime}$] & [km s$^{-1}$] & [pc] & [km s$^{-1}$] & [$10^{7}$ M$_{\odot}$] &  [$10^{7}$ M$_{\odot}$] \\
\hline
\endhead

\hline
\endfoot

 1 & 12 01 55.36 & -18 52 49 & 1414 &130 & 21 &  16 & 2.6 \\
 2 & 12 01 55.53 & -18 52 47 & 1514 &242 & 22 &  30 & 5.3 \\
 3 & 12 01 55.37 & -18 52 48 & 1374 &157 & 22 &  15 & 3.4 \\
 4 & 12 01 55.42 & -18 52 49 & 1504 &126 & 16 &  13 & 1.5 \\
 5 & 12 01 55.53 & -18 52 45 & 1364 &162 & 14 & 8.2 & 1.4 \\
 6 & 12 01 55.50 & -18 52 45 & 1394 &146 & 35 & 8.2 & 8.1 \\
 7 & 12 01 55.30 & -18 52 51 & 1464 &180 & 25 & 4.9 & 5.1 \\
 8 & 12 01 55.38 & -18 52 48 & 1464 &130 & 27 & 4.4 & 4.4 \\
 9 & 12 01 55.48 & -18 52 52 & 1474 &110 & 16 & 1.5 & 1.4 \\
10 & 12 01 55.38 & -18 52 51 & 1364 &151 & 18 & 2.4 & 2.3 \\
11 & 12 01 54.96 & -18 52 53 & 1464 &181 & 21 &  30 & 3.5 \\
12 & 12 01 54.74 & -18 52 52 & 1444 &205 & 24 &  24 & 5.4 \\
13 & 12 01 54.78 & -18 52 55 & 1604 &185 & 29 &  12 & 7.0 \\
14 & 12 01 54.78 & -18 52 53 & 1524 &116 & 28 & 5.0 & 4.0 \\
15 & 12 01 54.77 & -18 52 53 & 1544 &190 & 34 & 6.6 & 9.8 \\
16 & 12 01 54.77 & -18 52 57 & 1644 &  92 &  9 & 2.6 &0.34 \\
17 & 12 01 55.01 & -18 52 53 & 1564 &139 & 33 & 3.6 & 7.0 \\
18 & 12 01 55.12 & -18 52 54 & 1504 &153 & 22 & 7.7 & 3.2 \\
19 & 12 01 55.00 & -18 53 06 & 1494 &239 & 31 &  30 &  10 \\
20 & 12 01 54.92 & -18 53 03 & 1554 &208 & 28 &  15 & 7.2 \\
21 & 12 01 55.09 & -18 53 02 & 1514 &129 & 12 & 4.0 &0.87 \\
22 & 12 01 54.68 & -18 53 03 & 1454 &263 & 17 & 6.3 & 3.4 \\
23 & 12 01 54.89 & -18 53 00 & 1584 &143 & 11 & 3.5 &0.81 \\
24 & 12 01 54.92 & -18 53 00 & 1514 &137 & 20 & 5.1 & 2.5 \\
25 & 12 01 55.02 & -18 53 01 & 1504 &105 & 22 & 4.4 & 2.3 \\
26 & 12 01 54.96 & -18 53 08 & 1614 &243 & 13 & 6.0 & 1.9 \\
27 & 12 01 54.96 & -18 53 08 & 1584 &203 & 36 & 4.9 &  12 \\
28 & 12 01 54.66 & -18 53 05 & 1474 &226 & 31 & 7.3 & 9.9 \\
29 & 12 01 55.18 & -18 52 40 & 1574 &348 & 15 &  19 & 3.4 \\
30 & 12 01 54.66 & -18 52 45 & 1584 &251 & 21 & 7.8 & 4.9 \\
31 & 12 01 54.44 & -18 52 43 & 1614 &147 &  9 & 1.3 &0.57 \\
32 & 12 01 54.26 & -18 53 08 & 1414 &132 & 16 & 4.6 & 1.5 \\
33 & 12 01 54.33 & -18 53 08 & 1454 &203 & 15 & 7.1 & 2.1 \\
34 & 12 01 53.16 & -18 52 02 & 1624 &256 & 29 &  61 &  10 \\
35 & 12 01 53.22 & -18 52 04 & 1584 &199 & 24 &  34 & 5.0 \\
36 & 12 01 53.17 & -18 52 06 & 1624 &273 & 26 &  31 & 8.5 \\
37 & 12 01 53.49 & -18 52 06 & 1634 &218 & 25 & 4.6 & 6.0 \\
38 & 12 01 52.61 & -18 51 59 & 1714 & 93 & 10 & 1.6 &0.42 \\
39 & 12 01 53.26 & -18 52 04 & 1674 &225 & 22 & 7.6 & 4.8 \\
40 & 12 01 52.97 & -18 51 60 & 1634 &206 & 27 & 4.2 & 6.8 \\
41 & 12 01 53.18 & -18 52 06 & 1684 &217 & 12 & 3.0 & 1.4 \\
42 & 12 01 52.85 & -18 52 06 & 1634 &180 & 13 & 3.1 & 1.5 \\
43 & 12 01 52.76 & -18 52 07 & 1634 &  89 &  6 & 1.3 &0.13 \\
44 & 12 01 53.65 & -18 53 09 & 1604 &190 & 31 &  21 & 8.1 \\
45 & 12 01 53.32 & -18 53 09 & 1604 &107 & 31 &  12 & 4.6 \\
46 & 12 01 53.33 & -18 53 13 & 1624 &122 & 18 & 5.7 & 1.8 \\
47 & 12 01 53.65 & -18 53 11 & 1684 &130 & 28 & 7.4 & 4.5 \\
48 & 12 01 53.21 & -18 53 11 & 1644 &120 & 28 & 8.1 & 4.4 \\ 
49 & 12 01 53.45 & -18 53 09 & 1584 &178 & 33 &  10 & 8.6 \\
50 & 12 01 53.16 & -18 53 12 & 1704 &  85 & 16 & 2.6 & 1.0 \\
51 & 12 01 53.17 & -18 53 12 & 1684 &113 & 24 & 5.2 & 2.8 \\
52 & 12 01 53.25 & -18 53 18 & 1724 &142 & 20 & 2.2 & 2.6 \\
53 & 12 01 53.36 & -18 53 17 & 1704 &120 & 10 & 2.8 &0.55 \\
54 & 12 01 53.81 & -18 53 13 & 1644 &156 & 19 & 3.3 & 2.6 \\
55 & 12 01 53.34 & -18 53 11 & 1604 &119 & 17 & 3.1 & 1.5 \\
56 & 12 01 53.06 & -18 53 15 & 1704 &111 & 12 & 1.9 &0.77 \\
57 & 12 01 53.28 & -18 53 16 & 1684 &115 & 18 & 2.1 & 1.6 \\
\end{longtable}
\end{center}

\end{document}